\pdfoutput=1

\documentclass[reqno,12pt]{amsart}%
\usepackage{amsfonts}
\usepackage{amsmath}
\usepackage{amssymb}
\usepackage{graphicx}%
\providecommand{\U}[1]{\protect\rule{.1in}{.1in}}
\newtheorem{theorem}{Theorem}[section]

\newtheorem{corollary}[theorem]{Corollary}

\newtheorem{definition}[theorem]{Definition}
\newtheorem{example}[theorem]{Example}

\newtheorem{lemma}[theorem]{Lemma}

\newtheorem{proposition}[theorem]{Proposition}

\numberwithin{theorem}{section}

\begin{document}
\title[Analysis of operators]{Analysis of unbounded operators and random motion}
\author{Palle E.T. Jorgensen}
\address{Department of Mathematics\\
14 MLH\\
The University of Iowa\\
Iowa City, IA 52242-1419\\
USA}
\email{jorgen@math.uiowa.edu}
\urladdr{http://www.math.uiowa.edu}
\thanks{}
\thanks{Work supported in part by the US National Science Foundation.}
\date{}
\dedicatory{\textsc{University of Iowa}}
\begin{abstract}
We study infinite weighted graphs with view to \textquotedblleft limits at
infinity,\textquotedblright\ or boundaries at infinity.\ Examples of such
weighted graphs arise in infinite (in practice, that means \textquotedblleft
very\textquotedblright\ large) networks of resistors, or in statistical
mechanics models for classical or quantum systems. But more generally our
analysis includes reproducing kernel Hilbert spaces and associated operators
on them. If $X$ is some infinite set of vertices or nodes, in applications the
essential ingredient going into the definition is a reproducing kernel Hilbert
space; it measures the differences of functions on $X$ evaluated on pairs of
points in $X$. And the Hilbert norm-squared in $\mathcal{H}(X)$ will represent
a suitable measure of energy. Associated unbounded operators will define a
notion or dissipation, it can be a graph Laplacian, or a more abstract
unbounded Hermitian operator defined from the reproducing kernel Hilbert space
under study. We prove that there are two closed subspaces in reproducing
kernel Hilbert space $\mathcal{H}(X)$ which measure quantitative notions of
limits at infinity in $X$, one generalizes finite-energy harmonic functions in
$\mathcal{H}(X)$, and the other a deficiency index of a natural operator in
$\mathcal{H}(X)$ associated directly with the diffusion. We establish these
results in the abstract, and we offer examples and applications. Our results
are related to, but different from, potential theoretic notions of
\textquotedblleft boundaries\textquotedblright\ in more standard random walk
models. Comparisons are made.

\end{abstract}
\maketitle

\section{Introduction\label{Intro}}

We will use the theory of unbounded Hermitian operators with dense domain in
Hilbert space in a study of infinite weighted graphs with view to
\textquotedblleft limits at infinity.\textquotedblright\ We begin with
introducing the tools from operator theory as developed by H. M. Stone, John
von Neumann, Kurt Friedrichs, and Tosio Kato with view to our particular
setup. We stress that a Hermitian operator may not be selfadjoint, and that
the discrepancy is measured by deficiency-indices (details below and the books
\cite{St90} and \cite{vN32}, and more recently \cite{DuSc88} and
\cite{ReSi75}.) In physical problems, see e.g., \cite{Bal00}, these
mathematical notions of defect take the form of \textquotedblleft boundary
conditions;\textquotedblright\ for example waves that are diffracted on the
boundary of a region in Euclidean space; the scattering of classical waves on
a bounded obstacle \cite{LaPh80}; a quantum mechanical \textquotedblleft
particle\textquotedblright\ in a repulsive potential that shoots to infinity
in finite time; or in more recent applications (see e.g., \cite{JoPe08},
\cite{BBC+09}, \cite{BrWo05}, \cite{OrWo07}) random walk on infinite weighted
graphs $G$ that \textquotedblleft wander off\textquotedblright\ to points on
an idealized boundary of $G$. In all of the instances, one is faced with a
dynamical problem: For example, the solution to a Schr\"{o}dinger equation,
represents the time evolution of quantum states in a particular problem in
atomic physics.

The operators in these applications will be Hermitian, but in order to solve
the dynamical problems, one must first identify a selfadjoint extension of the
initially given operator. Once that is done, von Neumann's spectral theorem
can then be applied to the selfadjoint operator. A choice of selfadjoint
extension will have a spectral resolution, i.e., it is an integral of an
orthogonal projection valued measure; with the different extensions
representing different \textquotedblleft physical\textquotedblright\ boundary
conditions. Hence non-zero deficiency indices measure degrees of
non-selfadjointness, and deficiency spaces \textquotedblleft
measure\textquotedblright\ boundary obstructions or scattering on an obstacle.

The variety of applied problems that lend themselves to computation of
deficiency indices and the study of selfadjoint extensions are vast and
diverse. As a result, it helps if one can identify additional structures that
throw light on the problem.

Our results are inspired in part by the following recent developments in
related areas: fractals in the small and in the large \cite{DJ07a, DJ09,
JKS07}, representation theory \cite{DJ08a, DJ08b}, operator algebras
\cite{CJ08, So08}, harmonic analysis \cite{DJ07b, DJ09, DHP08, JKS08}; and
multiresolutions/wavelets \cite{AJP07, So06, JS07, JM08}.

We will further use notions from dynamics, infinite matrix products, to prove
essential selfadjointness of families of Hermitian operators arising naturally
in reproducing kernel Hilbert spaces. The latter include graph Laplacians for
infinite weighted graphs $(G,w)$ with the Laplacian in this context presented
as a Hermitian operator in an associated Hilbert space of finite energy
functions on the vertex set in $G$. Other examples include Hilbert spaces of
band-limited signals.

Further applications enter into the techniques used in discrete simulations of
stochastic integrals, see \cite{Hi80}.

We encountered the present operator theoretic results in our study of discrete
Laplacians, which in turn have part of its motivation in numerical analysis. A
key tool in applying numerical analysis to solving partial differential
equations is discretization, and use of repeated differences; see e.g.,
\cite{AH05}. Specifically, one picks a grid size $h$, and then proceeds in
steps: (1) Starting with a partial differential operator, then study an
associated discretized operator with the use of repeated differences on the
$h$-lattice in $\mathbb{R}^{d}$. (2) Solve the discretized problem for $h$
fixed. (3) As $h$ tends to zero, numerical analysts evaluate the resulting
approximation limits, and they bound the error terms. For this purpose, one
must use a metric, and the norm in Hilbert space has proved an effective tool,
hence the Hilbert spaces and the operator theory.

This procedure connects to our present graph-Laplacians: When discretization
is applied to the Laplace operator in $d$ continuous variables, the result is
the graph of integer points $\mathbb{Z}^{d}$ with constant weights. But if
numerical analysis is applied instead to a continuous Laplace operator on a
Riemannian manifold, the discretized Laplace operator will instead involve
infinite graph with variable weights, so with vertices in other configurations
than $\mathbb{Z}^{d}$.

Inside the technical sections we will use standard tools from analysis and
probability. References to the fundamentals include \cite{AH05}, \cite{Kol77},
\cite{Nel73} and \cite{YN08}.

There is a large literature covering the general theory of reproducing kernel
Hilbert spaces and its applications, see e.g., \cite{Alp06}, \cite{AL08a},
\cite{AL08b}, \cite{Aro50}, and \cite{Zh09}. Such applications include
potential theory, stochastic integration, and boundary value problems from
PDEs among others. In brief summary, a reproducing kernel Hilbert space
$\mathcal{H}$ consists of two things: a Hilbert space of functions $f$ on a
set $X$ , and a reproducing kernel $k$, i.e., a complex valued function $k$ on
$X\times X$ such that for every $x$ in $X$, the function $k(\cdot,x)$ is in
$\mathcal{H}$ and reproduces the value $f(x)$ from the inner product
$<k(\cdot,x),f>$ in $\mathcal{H}$, so the formula
\[
f(x)=<k(\cdot,x),f>
\]
holds for all $x$ in $X$ . Moreover, there is a set of axioms for a function
$k$ in two variables that characterizes precisely when it determines a
reproducing kernel Hilbert space. And conversely there are necessary and
sufficient conditions that apply to Hilbert spaces $\mathcal{H}$ and decide
when $\mathcal{H}$ is a reproducing kernel Hilbert space.

Here we shall restrict these \textquotedblleft reproducing\textquotedblright%
\ axioms and obtain instead a smaller class of reproducing kernel Hilbert
spaces. We add two additional axioms: Firstly, we will be reproducing not the
values themselves of the functions $f$ in $\mathcal{H}$, but rather the
differences $f(x)-f(y)$ for all pairs of points in $X$ ; and secondly we will
impose one additional axiom to the effect that the Dirac mass at $x$ is
contained in $\mathcal{H}$ for all $x$ in $\mathcal{H}$. When these two
additional conditions are satisfied, we say that $\mathcal{H}$ is a relative
reproducing kernel Hilbert space.

It is known that every weighted graph (the infinite case is of main concern
here) induces a relative reproducing kernel Hilbert space, and an associated
graph Laplacian. Under certain conditions, the converse holds as well: Given a
relative reproducing kernel Hilbert space $\mathcal{H}$ on a set $X$ , it is
then possible in a canonical way to construct a weighted graph $G$ such that
$X$ is the set of edges in $G$, and such that its energy Hilbert space
coincides with $\mathcal{H}$ itself. In our construction, the surprise is that
the edges in $G$ as well as the weights on the edges may be built directly
from only the Hilbert space axioms defining the initially given relative
reproducing kernel Hilbert space. Since this includes all infinite graphs of
electrical resistors and their potential theory (boundaries, harmonic
functions, and graph Laplacians) the result has applications to these fields,
and it serves to unify diverse branches in a vast research area.

\section{Operator Theoretic Framework\label{Frame}}

This section contains the precise definitions of the terms used above, and to
be used in later sections: the particular discrete networks, the weights on
edges, the associated Hilbert spaces, and the infinite Laplacians. We open
with two lemmas which establish links between the graph theoretic networks on
one side, and the operator theory (unbounded Hermitian operators) on the
other. This will allow us to encode certain \textquotedblleft
boundaries\textquotedblright\ with two subspaces of an associated Hilbert space.

Let $X$ be a set, and let $c:X\times X\rightarrow\mathbb{R}_{\geq0}$ a
function satisfying the following four conditions:

\begin{enumerate}
\item[(i)] For all $x\in X,$
\[
\#\left\{  y\in X|c\left(  x,y\right)  \not =0\right\}  <\infty\text{.}%
\]

\item[(ii)] Symmetry:
\[
c\left(  x,y\right)  =c\left(  y,x\right)  ,\forall x,y\in X\text{.}%
\]

\item[(iii)]
\[
c\left(  x,x\right)  =\sum_{y\in X}c\left(  x,y\right)  \text{.}%
\]

\item[(iv)] For all $x,y\in X$, such that $x\not =y$, there is a finite set
$\left\{  x_{0},x_{1},\ldots,x_{n}\right\}  $ of distinct points in $X$ such
that $c\left(  x_{i},x_{i+1}\right)  \not =0$, $0\leq i<n$, and $x_{0}=x$,
$x_{n}=y$.
\end{enumerate}

\begin{lemma}
\label{LemFrameA}Let $\left(  X,c\right)  $ be a system as described above
with the function
\[
c:X\times X\rightarrow\mathbb{R}_{\geq0}%
\]
satisfying \emph{(}i\emph{)}--\emph{(}iv\emph{)}.

For $x\in X$, let $\delta_{x}$ be the Dirac-function on $X$ supported at $x$.

Then there is a Hilbert space $\mathcal{H}\left(  X,c\right)  $ containing
$\left\{  \delta_{x}\right\}  _{x\in X}$ with inner product $\left\langle
\cdot,\cdot\right\rangle $ such that
\begin{equation}
\left\langle \delta_{x},\delta_{x}\right\rangle =c\left(  x,x\right)  \text{,}
\label{EqFrame0a}%
\end{equation}
and
\begin{equation}
\left\langle \delta_{x},\delta_{y}\right\rangle =-c\left(  x,y\right)  \text{
if }x\not =y\text{.} \label{EqFrame0b}%
\end{equation}

\end{lemma}

\begin{proof}
We will be working with functions on $X$ modulo multiples of the constant
function $1\!\!1$ on $X$. We define the Hilbert space $\mathcal{H}\left(
X,c\right)  $ as follows:
\begin{equation}
\mathcal{H}\left(  X,c\right)  \text{:}=\{\text{all functions }f:X\rightarrow
C\!\!\!\!C\text{ s.t. }\nonumber
\end{equation}%
\begin{equation}
\underset{%
\genfrac{}{}{0pt}{}{x~~y}{\text{s.t. }x\not =y}%
}{\sum\sum}c\left(  x,y\right)  \left\vert f\left(  x\right)  -f\left(
y\right)  \right\vert ^{2}<\infty\}; \label{EqFrame0c}%
\end{equation}
and we set
\begin{align}
\left\Vert f\right\Vert ^{2}  &  =\left\langle f,f\right\rangle \nonumber\\
&  =\mathcal{E}\left(  f\right)  \text{:}=\frac{1}{2}\underset{x\not =y}%
{\sum\sum}c\left(  x,y\right)  \left\vert f\left(  x\right)  -f\left(
y\right)  \right\vert ^{2}\text{ for }f\in\mathcal{H}\left(  X,c\right)
\text{.} \label{EqFrame0d}%
\end{align}

It is immediate that $\mathcal{H}\left(  X,c\right)  $ is then a Hilbert
space, and that the Dirac-functions $\delta_{x}$ are in $\mathcal{H}\left(
X,c\right)  $. If $\left\langle \cdot,\cdot\right\rangle $ is the inner
product corresponding to (\ref{EqFrame0d}), then a direct computation shows
that (\ref{EqFrame0a}) and (\ref{EqFrame0b}) are satisfied.
\end{proof}

\begin{lemma}
\label{LemFrameB}Let $\left(  X,c\right)  $ be a system satisfying conditions
\emph{(}i\emph{)}--\emph{(}iv\emph{)} above, and let $\mathcal{H}\left(
X,c\right)  ,\left\langle \cdot,\cdot\right\rangle $ be the Hilbert space
introduced in Lemma \ref{LemFrameA}.

For elements $u\in\mathcal{H}\left(  X,c\right)  $, set
\begin{equation}
\left(  \Delta u\right)  \left(  x\right)  \text{\emph{:}}=\left\langle
\delta_{x},u\right\rangle \text{ for }x\in X\text{.} \label{EqFrame0e}%
\end{equation}
Then $\Delta$ is a Hermitian operator with a dense domain $\mathcal{D}$ in
$\mathcal{H}$\emph{:}$=\mathcal{H}\left(  X,c\right)  $.
\end{lemma}

\begin{proof}
Select some base-point $o$ in $X$. Let $x\in X$, and select $\left(
x_{i}\right)  _{i=0}^{n}\subset X$ such that $x_{i-1}\not =x_{i}$, $c\left(
x_{i-1},x_{i}\right)  \not =0$, and $x_{0}=0$, $x_{n}=x$. Then if
$f\in\mathcal{H}$, we get the following estimate
\begin{equation}
\left\vert f\left(  x\right)  -f\left(  o\right)  \right\vert \leq\left(
\sum_{i=1}^{n}\frac{1}{c\left(  x_{i-1},x_{i}\right)  }\right)  ^{\frac{1}{2}%
}\left\Vert f\right\Vert \label{EqFrame0f}%
\end{equation}
where $\left\Vert f\right\Vert $ is the norm in $\mathcal{H}$. By Riesz'
lemma, there is a unique $v_{x}\in\mathcal{H}$ such that
\begin{equation}
f\left(  x\right)  -f\left(  o\right)  =\left\langle v_{x},f\right\rangle
\text{ for all }f\in\mathcal{H}\text{.} \label{EqFrame0g}%
\end{equation}

Now let $\mathcal{D}$ be the linear span of $\left(  v_{x}\right)  _{x\in X}$.
It follows from (\ref{EqFrame0g}) that $\mathcal{D}$ is dense in $\mathcal{H}%
$, and that
\begin{equation}
\Delta v_{x}=\delta_{x}-\delta_{o}\text{.} \label{EqFrame0h}%
\end{equation}

To prove (\ref{EqFrame0h}) first note that the functions $\left(
v_{x}\right)  $ in (\ref{EqFrame0g}) must be real-valued. This is a
consequence of the uniqueness in Riesz. We now verify (\ref{EqFrame0h}) by the
following computation:
\begin{align*}
\left(  \Delta v_{x}\right)  \left(  y\right)   &  =_{\left(  \text{by
(\ref{EqFrame0e})}\right)  }\left\langle \delta_{y},v_{x}\right\rangle \\
&  =_{\left(  \text{by (\ref{EqFrame0g})}\right)  }\delta_{y}\left(  x\right)
-\delta_{y}\left(  o\right) \\
&  =\left(  \delta_{x}-\delta_{o}\right)  \left(  y\right)  \text{,}%
\end{align*}
which is the desired conclusion (\ref{EqFrame0h}).

A direct inspection yields the following two additional properties.
\begin{align*}
\left\langle \Delta u_{1},u_{2}\right\rangle  &  =\left\langle u_{1},\Delta
u_{2}\right\rangle \text{, and}\\
\left\langle u,\Delta u\right\rangle  &  \geq0\text{, for all }u_{1}%
,u_{2}\text{, and }u\in\mathcal{D}\text{.}%
\end{align*}

\end{proof}

\begin{definition}
\label{DefFrame1}~\newline

\begin{enumerate}
\item[(a)] Let $G^{0}$ and $G^{1}$ be two sets (in our case the infinite cases
are of main importance), and assume the following:

\begin{itemize}
\item $G^{1}\subset G^{0}\times G^{0}$.

\item If $\left(  x,y\right)  \in G^{1}$, then $x\not =y$, and we write $x\sim
y$.

\item If $\left(  x,y\right)  \in G^{1}$ then $\left(  y,x\right)  \in G^{1}$.

\item For every pair of points $x$ and $y$ in $G^{0}$, there is a
\emph{finite} subset
\[
\{\left(  x_{i-1},x_{i}\right)  \,|\,i=1,2,\cdots,n\}\subset G^{1}%
\]
such that $x_{0}=x$ and $x_{n}=y$.

\item For every $x\in G^{0}$ we assume that the set of neighbors is finite,
i.e.,%
\begin{equation}
\#\operatorname*{Nbh}\nolimits_{G}\left(  x\right)  <\infty, \label{EqFrame1}%
\end{equation}
where
\begin{equation}
\#\operatorname*{Nbh}\nolimits_{G}\left(  x\right)  =\left\{  y\in G^{0}|y\sim
x\right\}  \text{.} \label{EqFrame2}%
\end{equation}

\end{itemize}

\item[(b)] Let $c$:$~G^{1}\rightarrow\mathbb{R}_{+}$ be a function satisfying
\begin{equation}
c\left(  x,y\right)  =c\left(  y,x\right)  \text{ for all }\left(  x,y\right)
\in G^{1}\text{.} \label{EqFrame3}%
\end{equation}

\item[(c)] For functions $u$:$~G^{0}\rightarrow\mathbb{C}$ set
\begin{equation}
\mathcal{E}\left(  u\right)  \text{\emph{:}}\underset{%
\genfrac{}{}{0pt}{}{\text{all }xy}{\text{such that }x\sim y}%
}{=\frac{1}{2}\sum\sum}c\left(  x,y\right)  \left\vert u\left(  x\right)
-u\left(  y\right)  \right\vert ^{2}\text{.} \label{EqFrame4}%
\end{equation}

\end{enumerate}
\end{definition}

We will consider the Hilbert space $\mathcal{H}_{E}$ of all functions modulo
constants such that
\begin{equation}
\mathcal{E}\left(  u\right)  <\infty\text{;} \label{EqFrame5}%
\end{equation}
called the \textit{energy Hilbert space}, and (\ref{EqFrame5}) referring to
\textit{finite energy}.

For functions $u$:$~G^{0}\rightarrow\mathbb{C}$, we define the \textit{graph
Laplacian}, or simply the \textit{Laplace operator} $\Delta$ by
\begin{equation}
\left(  \Delta u\right)  \left(  x\right)  =\sum_{y\sim x}c\left(  x,y\right)
\left(  u\left(  x\right)  -u\left(  y\right)  \right)  \text{.}
\label{EqFrame6}%
\end{equation}

\begin{proposition}
\label{PropFrame1}\ \newline

\begin{enumerate}
\item[(a)] The Laplace operator $\Delta$ in \emph{(\ref{EqFrame6})} is defined
on a dense linear subspace $\mathcal{D}$ in $\mathcal{H}_{E}$, and $\Delta$
maps $\mathcal{D}$ into $\mathcal{H}_{E}$.

\item[(b)] Pick some $o$ in $G^{0}$. Then for every $x\in G^{0}\diagdown
\left(  0\right)  $, there is a unique $v_{x}\in\mathcal{H}_{E}$ such that
\begin{equation}
\left\langle v_{x},u\right\rangle _{E}=u\left(  x\right)  -u\left(  o\right)
\text{ for all }u\in\mathcal{H}_{E}\text{.} \label{EqFrame7}%
\end{equation}

\item[(c)] The function $v_{x}$ in \emph{(}\ref{EqFrame7}\emph{)} satisfies
\begin{equation}
\Delta v_{x}=\delta_{x}-\delta_{o}\text{.} \label{EqFrame8}%
\end{equation}

\item[(d)] For the subspace $\mathcal{D}$ in \emph{(}a\emph{)} we may take
\begin{equation}
\mathcal{D}=\operatorname*{span}\left\{  v_{x}|x\in G^{0}\diagdown\left(
o\right)  \right\}  \text{,} \label{EqFrame9}%
\end{equation}
i.e., all finite linear combinations of the $v_{x}$ family of vectors.

\item[(e)] The Dirac-functions $\delta_{x}$ in \emph{(}\ref{EqFrame8}\emph{)}
satisfy
\begin{equation}
\delta_{x}=c\left(  x\right)  v_{x}-\sum_{y\sim x}c\left(  x,y\right)  v_{y}
\label{EqFrame10}%
\end{equation}
where
\begin{equation}
c\left(  x\right)  \text{\emph{:}}=\sum_{y\sim x}c\left(  x,y\right)  \text{.}
\label{EqFrame11}%
\end{equation}

\item[(f)] The operator $\Delta$ is \emph{Hermitian} on $\mathcal{D}$, i.e.,
\begin{equation}
\left\langle \Delta u_{1},u_{2}\right\rangle =\left\langle u_{1},\Delta
u_{2}\right\rangle \text{ holds for all }u_{1},u_{2}\in\mathcal{D}\text{,}
\label{EqFrame12}%
\end{equation}
where
\begin{align}
\left\langle u_{1},u_{2}\right\rangle \text{{}}  &  \text{\emph{:}%
}=\left\langle u_{1},u_{2}\right\rangle _{E}\nonumber\\
\text{{}}  &  \text{\emph{:}}=\mathcal{E}\left(  u_{1},u_{2}\right)
\nonumber\\
\text{{}}  &  \text{\emph{:}}=\underset{%
\genfrac{}{}{0pt}{}{x\quad\,y}{\text{such that }x\sim y}%
}{\frac{1}{2}\sum\sum}c\left(  x,y\right)  \left(  \overline{u_{1}\left(
x\right)  }-\overline{u_{1}\left(  y\right)  }\right)  \left(  u_{2}\left(
x\right)  -u_{2}\left(  y\right)  \right)  \text{.} \label{EqFrame13}%
\end{align}

\item[(g)] Semiboundedness\emph{:}
\begin{equation}
\left\langle u,\Delta u\right\rangle \geq0\text{ for all }u\in\mathcal{D}%
\text{.} \label{EqFrame14}%
\end{equation}

\end{enumerate}
\end{proposition}

\begin{proof}
~\newline

\begin{enumerate}
\item[(a)] Follows from (b) and (c).

\item[(b)] This is an application of Riesz' lemma for the Hilbert space
$\mathcal{H}_{E}$; see \cite{JoPe08}. Indeed, pick $\left(  x_{i}\right)
_{i=0}^{n}\subset G^{0}$ such that $\left(  x_{i-1},x_{i}\right)  \in G^{1}$,
and $x_{0}=0$, $x_{n}=x$; then
\begin{equation}
\left\vert u\left(  x\right)  -u\left(  o\right)  \right\vert \leq\left(
\sum_{i=1}^{n}\frac{1}{c\left(  x_{i-1},x_{i}\right)  }\right)  ^{\frac{1}{2}%
}\left\Vert u\right\Vert _{\mathcal{H}_{E}} \label{EqFrame15}%
\end{equation}
where
\begin{equation}
\left\Vert u\right\Vert _{\mathcal{H}_{E}}\text{:}=\mathcal{E}\left(
u\right)  ^{\frac{1}{2}}\text{, }u\in\mathcal{H}_{E}\text{.} \label{EqFrame16}%
\end{equation}
The desired conclusion (\ref{EqFrame7}) is then immediate from Riesz.

\item[(c)] It is clear from (\ref{EqFrame7}) that each function
\begin{equation}
v_{x}\text{:}~G^{0}\rightarrow\mathbb{R} \label{EqFrame17}%
\end{equation}
is real valued and that $\{v_{x}|x\in G^{0}\diagdown\left(  0\right)  \}$ span
a dense subspace $\mathcal{D}$ in $\mathcal{H}_{E}$. \newline

Hence to prove (\ref{EqFrame8}), we need to show that
\begin{equation}
\left\langle v_{y},\Delta v_{x}-\left(  \delta_{x}-\delta_{0}\right)
\right\rangle =0\text{ for all }x,y\in G^{0}\diagdown\left(  0\right)
\text{.} \label{EqFrame18}%
\end{equation}
~\newline

Because of (\ref{EqFrame4}), we may impose the following re-\linebreak
normalization:
\begin{equation}
v_{x}\left(  o\right)  =0,~\forall x\in G^{0}\diagdown\left(  0\right)
\text{.} \label{EqFrame19}%
\end{equation}
With this, now (\ref{EqFrame18}) follows from (\ref{EqFrame4}),
(\ref{EqFrame6}) and (\ref{EqFrame7}), by a direct computation which we leave
to the reader; see also \cite{JoPe08}.

\item[(d)] Follows from (c).

\item[(e)] By (d) it is enough to prove that for all $z\in G^{0}$, we have the
identity:
\[
\left\langle v_{z},\delta_{x}-c\left(  x\right)  v_{x}+\sum_{y\sim x}c\left(
x,y\right)  v_{y}\right\rangle =0\text{,}%
\]
which in turn is a computation similar to the one used in (c).

\item[(f)] and (g) Follow by polarization, and a direct computation of
(\ref{EqFrame14}). Indeed, if
\begin{equation}
u=\sum_{x}\xi_{x}v_{x} \label{EqFrame20}%
\end{equation}
is a finite summation over $x\in G^{0}\diagdown\left(  0\right)  $, $\xi
_{x}\in\mathbb{C}$, then $\xi_{x}=\left(  \Delta u\right)  \left(  x\right)
$, and
\begin{equation}
\left\langle u,\Delta u\right\rangle =\sum_{x}\left\vert \Delta u\left(
x\right)  \right\vert ^{2}+\left\vert \sum_{x}\left(  \Delta u\right)  \left(
x\right)  \right\vert ^{2}\text{.} \label{EqFrame21}%
\end{equation}

\end{enumerate}
\end{proof}

\begin{definition}
\label{DefFrame2}The following two closed subspaces play a critical role in
our understanding of \emph{geometric boundaries} of weighted graphs $\left(
G,c\right)  $\emph{;} i.e., a graph with $G^{0}$ serving as the set of
\emph{vertices}, $G^{1}$ the set of \emph{edges}, and
\begin{equation}
c\text{\emph{:~}}G^{1}\rightarrow\mathbb{R}_{+} \label{EqFrame22}%
\end{equation}
a fixed \emph{conductance function}.
\end{definition}

Given $\left(  G,c\right)  $ we introduce the Hilbert space $\mathcal{H}_{E}$
and the operator $\Delta$ as in (\ref{EqFrame4}), (\ref{EqFrame5}) and
(\ref{EqFrame6}) above. Note that both depend on the choice of the function
$c$ in (\ref{EqFrame22}).

\begin{enumerate}
\item[(a)] Set
\begin{equation}
\operatorname*{Harm}\text{:}=\left\{  h\in\mathcal{H}_{E}|\Delta h=0\right\}
\text{;} \label{EqFrame23}%
\end{equation}
the finite-energy harmonic functions, \textquotedblleft Harm\textquotedblright%
\ is short for \textquotedblleft harmonic\textquotedblright. Note that
functions $h$ in $\operatorname*{Harm}$ may be unbounded as functions on
$G^{0}$. ~\newline

Setting $\operatorname*{Fin}$:$=$ the closed span of $\left\{  \delta_{x}|x\in
G^{0}\right\}  $ in $\mathcal{H}_{E}$, it is easy to see that
\begin{equation}
\operatorname*{Harm}=\mathcal{H}_{E}\ominus\operatorname*{Fin}\text{,}
\label{EqFrame24}%
\end{equation}
and
\begin{equation}
\operatorname*{Fin}=\mathcal{H}_{E}\ominus\operatorname*{Harm}\text{,}
\label{EqFrame25}%
\end{equation}
i.e., that the decomposition
\begin{equation}
\mathcal{H}_{E}\text{:}=\operatorname*{Fin}\oplus\operatorname*{Harm}
\label{EqFrame26}%
\end{equation}
holds; see \cite{JoPe08}.

\item[(b)] Set
\begin{equation}
\operatorname*{Def}\text{:}=\left\{  u\in\mathcal{H}_{E}|\Delta u=-u\right\}
\text{;} \label{EqFrame27}%
\end{equation}
the finite-energy deficiency vectors for the operator $\Delta$. ~\newline

(The reader will notice that solutions $u\not =0$ to (\ref{EqFrame27}) appear
to contradict the estimates (\ref{EqFrame14}) and (\ref{EqFrame21}); but this
is not so. The deeper explanation lies in the theory of
\textit{deficiency-indices}, and \textit{deficiency-subspaces} for unbounded
linear operators in Hilbert space.) The notation \textquotedblleft
Def\textquotedblright\ is short for deficiency-subspace.
\end{enumerate}

\section{Computations\label{Compute}}

While the analysis of the two closed subspaces from Definition \ref{DefFrame2}
(the finite energy harmonic functions, and the functions in the defect-space,
or deficiency-space, for the Laplacian) will be carried out in general
(sections \ref{Space} and \ref{Embed}), it will be useful to work them out in
a particular family of special cases; examples. At a first glance, these
examples may indeed appear rather special, but we will show in section
\ref{Embed} that they have wider use in our analysis of the \textquotedblleft
boundary at infinity.\textquotedblright\ The special cases discussed here will
further show that the abstract spaces and operators from section \ref{Frame}
allow for explicit computations.

\begin{example}
\label{ExCompute1}In an infinite weighted graph $\left(  G,c\right)  $ take
for vertex-set
\begin{equation}
G^{0}=\left\{  0\right\}  \cup\mathbb{Z}_{+}=\left\{  0,1,2,\cdots\right\}
\text{,} \label{EqCompute1}%
\end{equation}
edges, nearest neighbors,
\begin{equation}
\operatorname*{Nbh}\nolimits_{G}\left(  0\right)  =\left\{  1\right\}
\text{,} \label{EqCompute2}%
\end{equation}
and
\begin{equation}
\operatorname*{Nbh}\nolimits_{G}\left(  x\right)  =\left\{  x-1,x+1\right\}
\text{ if }x\in\mathbb{Z}_{+}\text{.} \label{EqCompute3}%
\end{equation}
Pick a function $\mu$ on $\mathbb{Z}_{+}$, and set
\begin{equation}
c\left(  x-1,x\right)  =\text{\emph{:}}\mu\left(  x\right)  \text{ if }%
x\in\mathbb{Z}_{+}\text{.} \label{EqCompute4}%
\end{equation}

\end{example}

We will be interested in an \textit{extended} version of the example when
$G^{0}=\mathbb{Z}$, and
\begin{equation}
\operatorname*{Nbh}\nolimits_{G}\left(  x\right)  =\left\{  x-1,x+1\right\}
\label{EqCompute5}%
\end{equation}
for all $x\in\mathbb{Z}$. In this case, we will extend (\ref{EqCompute4}) by
symmetry, as follows:
\begin{equation}
c\left(  x-1,x\right)  =\text{:}\mu\left(  x\right)  \text{ if }x\in
\mathbb{Z}_{+}\text{,} \label{EqCompute6}%
\end{equation}
and
\begin{equation}
c\left(  x,x+1\right)  =\mu\left(  \left\vert x\right\vert \right)  \text{ if
}x\in\mathbb{Z}_{-}\text{.} \label{EqCompute7}%
\end{equation}

To distinguish the two cases we will denote the first one $\left(  G_{+}%
,\mu\right)  $, and the second $\left(  G,\mu\right)  $.

It turns out that there are important distinctions between the two, relative
to the two subspaces $\operatorname*{Harm}$ and $\operatorname*{Def}$
introduced above.

The matrix form of the operator $\Delta$ is as follows in the two cases:
\smallskip%

\begin{table}[h] \centering
$\left(
\begin{tabular}
[c]{lllllll}%
$\scriptstyle\mu\left(  1\right)  $ & $\scriptstyle-\mu\left(  1\right)  $ &
$\scriptstyle0$ & $\scriptstyle0$ & $\scriptstyle0$ & $\cdots$ &
$\scriptstyle\rightarrow\infty$\\
$\scriptstyle-\mu\left(  1\right)  $ & $\scriptstyle\mu\left(  1\right)
+\mu\left(  2\right)  $ & $\scriptstyle-\mu\left(  2\right)  $ &
$\scriptstyle0$ & $\scriptstyle0$ &  & \\
$\scriptstyle0$ & $\scriptstyle-\mu\left(  2\right)  $ & $\scriptstyle\mu
\left(  2\right)  +\mu\left(  3\right)  $ & $\scriptstyle-\mu\left(  3\right)
$ & $\scriptstyle0$ &  & \\
$\scriptstyle0$ & $\scriptstyle0$ & $\scriptstyle-\mu\left(  3\right)  $ &
$\scriptstyle\mu\left(  3\right)  +\mu\left(  4\right)  $ & $\scriptstyle-\mu
\left(  4\right)  $ &  & \\
$\scriptstyle0$ & $\scriptstyle0$ & $\scriptstyle0$ & $\scriptstyle-\mu\left(
4\right)  $ & $\scriptstyle\mu\left(  4\right)  +\mu\left(  5\right)  $ &
$\cdots$ & \\
$\scriptstyle\vdots$ & $\scriptstyle\vdots$ & $\scriptstyle\vdots$ &  &  &
$\cdots$ & \\
$\underset{\infty}{\downarrow}$ &  &  &  &  &  &
\end{tabular}
\right)  \medskip$%
\caption{\bfseries{Matrix of \boldmath{$\Delta$} in \boldmath{$(G_+,\mu)$}}}\label{TableKey copy(2)}%
\end{table}%

\begin{center}%
\begin{table}[h]\centering
$\!\left(  \!%
\begin{tabular}
[c]{lccc|c|ccc}
& $\scriptstyle0$ &  &  &  &  &  & \\
& $\scriptstyle0$ & $\scriptstyle0$ & $\scriptstyle0$ & $\cdots$ & $\cdots$ &
& \\
$\scriptstyle\ddots$ & $\scriptstyle-\mu\left(  4\right)  $ & $\scriptstyle0$
& $\scriptstyle0$ & $\scriptstyle0$ & $\cdots$ & $\cdots$ & \\
& $\scriptstyle\mu(3)+\mu(4)$ & $\scriptstyle-\mu\left(  3\right)  $ &
$\scriptstyle0$ & $\scriptstyle0$ & $\scriptstyle0$ & $\cdots$ & \\
& $\scriptstyle-\mu\left(  3\right)  $ & $\scriptstyle\mu(2)+\mu(3)$ &
$\scriptstyle-\mu\left(  2\right)  $ & $\scriptstyle0$ & $\scriptstyle0$ &
$\scriptstyle0$ & $\cdots$\\
& $\scriptstyle0$ & $\scriptstyle-\mu\left(  2\right)  $ & $\scriptstyle\mu
(1)+\mu(2)$ & $\scriptstyle-\mu\left(  1\right)  $ & $\scriptstyle0$ &
$\scriptstyle0$ & $\scriptstyle0$\\\hline
$\cdots$ & $\scriptstyle0$ & $\scriptstyle0$ & $\scriptstyle-\mu\left(
1\right)  $ & $\scriptstyle2\mu\left(  1\right)  $ & $\scriptstyle-\mu\left(
1\right)  $ & $\scriptstyle0$ & $\scriptstyle0$\\\hline
$\cdots$ & $\scriptstyle0$ & $\scriptstyle0$ & $\scriptstyle0$ &
$\scriptstyle-\mu\left(  1\right)  $ & $\scriptstyle\mu\left(  1\right)
+\mu\left(  2\right)  $ & $\scriptstyle-\mu\left(  2\right)  $ &
$\scriptstyle0$\\
& $\cdots$ & $\cdots$ & $\scriptstyle0$ & $\scriptstyle0$ & $\scriptstyle-\mu
\left(  2\right)  $ & $\scriptstyle\mu\left(  2\right)  +\mu\left(  3\right)
$ & $\scriptstyle-\mu\left(  3\right)  $\\
&  &  & $\cdots$ & $\cdots$ & $\scriptstyle0$ & $\scriptstyle-\mu\left(
3\right)  $ & $\scriptstyle\mu\left(  3\right)  +\mu\underset{\ddots}{\left(
4\right)  }$\\
&  &  &  &  &  & $\scriptstyle0$ & $\scriptstyle0$%
\end{tabular}
\!\!\right)  \bigskip$\label{TableKey}%
\caption{\bfseries{Matrix of \boldmath{$\Delta$} in \boldmath{$(G,\mu)$}}}%
\end{table}%

\end{center}

Let $M\in\mathbb{R}_{+}$; suppose $M>1$, and set $m=1/M$. For the conductance
function $\mu$ in (\ref{EqCompute6}) and (\ref{EqCompute7}) we set
\begin{equation}
\mu\left(  x\right)  \text{:}=M^{x}=\exp\left(  x\ln M\right)  \text{, }%
x\in\mathbb{Z}_{+}\text{.} \label{EqCompute8}%
\end{equation}

\begin{theorem}
\label{TheoCompute1}In the examples with $\mu$ as in \emph{(}\ref{EqCompute8}%
\emph{)}, we have the following%

\begin{table}[h] \centering
\begin{tabular}
[c]{ccc}\hline
\multicolumn{1}{|c}{} & \multicolumn{1}{|c}{$\left(  G_{+},\mu\right)  $} &
\multicolumn{1}{|c|}{$\left(  G,\mu\right)  $}\\\hline
\multicolumn{1}{|l}{\upshape{Harm}} & \multicolumn{1}{|l}{$^{1}$
{\small \upshape{zero}}} & \multicolumn{1}{|l|}{$^{2}$
\upshape{one-dimensional}}\\\hline
\multicolumn{1}{|l}{\upshape{Def}} & \multicolumn{1}{|l}{$^{3}$
\upshape{one-dimensional}} & \multicolumn{1}{|l|}{$^{4}$
\upshape{dimension zero or one; see details below!}}\\\hline
&  &
\end{tabular}
\caption{The four possibilities marked in the table will be proved in details below in parts 1 through 4.}\label{TableKey copy(1)}%
\end{table}%

\end{theorem}

\begin{proof}
It will be convenient for us to organize the proof into the four parts
indicated in Table 3.

\medskip

\textbf{Part 1.} With the definitions as given, we consider the possible
solutions $h$:$\left\{  0\right\}  \cup\mathbb{Z}_{+}\rightarrow\mathbb{C}$ to
the equation
\begin{equation}
\Delta h=0\text{.} \label{EqCompute9}%
\end{equation}
Hence
\begin{equation}
0=\mu\left(  1\right)  \left(  h\left(  0\right)  -h\left(  1\right)  \right)
\text{, } \label{EqCompute10}%
\end{equation}
and
\begin{equation}
0=\mu\left(  x\right)  \left(  h\left(  x\right)  -h\left(  x-1\right)
\right)  +\mu\left(  x+1\right)  \left(  h\left(  x\right)  -h\left(
x+1\right)  \right)  \text{ for }x\in\mathbb{Z}_{+}\text{.}
\label{EqCompute11}%
\end{equation}
Setting
\begin{equation}
\left(  \delta h\right)  \left(  x\right)  \text{:}=h\left(  x\right)
-h\left(  x-1\right)  \text{,} \label{EqCompute12}%
\end{equation}
equations (\ref{EqCompute10}) and (\ref{EqCompute11}) then take the following
form:
\begin{equation}
\mu\left(  1\right)  \left(  \delta h\right)  \left(  1\right)  =0\text{,}
\label{EqCompute13}%
\end{equation}
and
\begin{equation}
\mu\left(  x\right)  \left(  \delta h\right)  \left(  x\right)  =\mu\left(
x+1\right)  \left(  \delta h\right)  \left(  x+1\right)  \text{ for all }%
x\in\mathbb{Z}_{+}\text{.} \label{EqCompute14}%
\end{equation}
Since $\mu>0$, clearly the last two equations imply $\delta h\equiv0$. So
\[
h\left(  x\right)  =h\left(  0\right)  +\delta\left(  1\right)  +\cdots
+\left(  \delta h\right)  \left(  x\right)  =h\left(  0\right)  \text{,}%
\]
and $h$ must be the constant function. But $\mathcal{H}_{E}$ is obtained by
modding out with the constants, so $h=0$ in $\mathcal{H}_{E}$.

\medskip

\textbf{Part 2.} In this case, $G^{0}=\mathbb{Z}$, and we set $h\left(
0\right)  =0$, and
\begin{equation}
h\left(  -x\right)  =-h\left(  x\right)  \text{, }x\in\mathbb{Z}\text{.}
\label{EqCompute15}%
\end{equation}

In that case,
\begin{align*}
0  &  =\left(  \Delta h\right)  \left(  0\right)  =\mu\left(  1\right)
\left(  h\left(  0\right)  -h\left(  -1\right)  +h\left(  0\right)  -h\left(
1\right)  \right) \\
&  =\mu\left(  1\right)  \left(  2h\left(  0\right)  -h\left(  -1\right)
-h\left(  1\right)  \right)  \text{,}%
\end{align*}
so conditions (\ref{EqCompute15}) work. This together with (\ref{EqCompute14})
then determine $h$ on $\mathbb{Z}$.

Pick some constant $t\in\mathbb{R}_{+}$, and set
\begin{equation}
\mu\left(  x\right)  \delta h\left(  x\right)  =t\text{ for all }%
x\in\mathbb{Z}_{+}\text{;} \label{EqCompute16}%
\end{equation}
and then
\begin{align*}
h\left(  x\right)   &  =\delta h\left(  1\right)  +\delta h\left(  2\right)
+\cdots+\left(  \delta h\right)  \left(  x\right) \\
&  =t\left(  \mu\left(  1\right)  ^{-1}+\mu\left(  2\right)  ^{-1}+\cdots
+\mu\left(  x\right)  ^{-1}\right)  \text{.}%
\end{align*}
With conductance $\mu\left(  x\right)  =M^{x}$, and resistance $\xi$:$=M^{-1}%
$, we get
\begin{align}
h\left(  x\right)   &  =t\left(  \xi+\xi^{2}+\cdots+\xi^{x}\right)
\label{EqCompute17}\\
&  =t\xi\frac{1-\xi^{x}}{1-\xi}\underset{x\rightarrow\infty}{\longrightarrow
}\frac{t\xi}{1-\xi}\text{.}\nonumber
\end{align}

Moreover,
\begin{align*}
\mathcal{E}\left(  h\right)   &  =2\sum_{x=1}^{\infty}\mu\left(  x\right)
\left(  \left(  \delta h\right)  \left(  x\right)  \right)  ^{2}\\
&  =2\sum_{x=1}^{\infty}M^{x}\left(  t\xi^{x}\right)  ^{2}\\
&  =2t^{2}\sum_{x=1}^{\infty}\xi^{x}=\frac{2t^{2}\xi}{1-\xi}<\infty\text{.}%
\end{align*}

\end{proof}

\medskip

\textbf{Part 3.} Now the \textquotedblleft defect equation\textquotedblright%
\ is
\begin{equation}
\Delta u=-u,~u\in\mathcal{H}_{E}\text{,} \label{EqCompute18}%
\end{equation}
starting with
\[
\left(  \Delta u\right)  \left(  0\right)  =\mu\left(  1\right)  \left(
u\left(  0\right)  -u\left(  1\right)  \right)  =-u\left(  0\right)  \text{;}%
\]
so
\[
\left(  \delta u\right)  \left(  1\right)  =u\left(  1\right)  -u\left(
0\right)  =\xi u\left(  0\right)  \text{,}%
\]
or
\begin{equation}
u\left(  1\right)  =\left(  1+\xi\right)  u\left(  0\right)  \text{.}
\label{EqCompute19}%
\end{equation}

Equation (\ref{EqCompute18}) for $x\in\mathbb{Z}_{+}$ yields
\begin{equation}
\mu\left(  x+1\right)  \left(  \delta u\right)  \left(  x+1\right)
=\mu\left(  x\right)  \left(  \delta u\right)  \left(  x\right)  +u\left(
x\right)  \text{.} \label{EqCompute20}%
\end{equation}
Now set $u\left(  0\right)  =\lambda$. For $x=n\in\mathbb{Z}_{+}$, we now
define two sequences of polynomials as follows
\begin{equation}
\left(  \delta u\right)  \left(  n\right)  \text{:}=\xi^{n}p_{n}\left(
\xi\right)  \lambda\text{,} \label{EqCompute21}%
\end{equation}%
\begin{equation}
u\left(  n\right)  \text{:}=q_{n}\left(  \xi\right)  \lambda\text{,}
\label{EqCompute22}%
\end{equation}
where
\begin{equation}
\xi=M^{-1}\text{, and }\mu\left(  n\right)  =M^{n}\text{.} \label{EqCompute23}%
\end{equation}

It follows that $u$ is a multiple of the function%
\begin{equation}
u_{1}\left(  x\right)  \text{:}=q_{x}\left(  \xi\right)  \text{, }x\in
0\cup\mathbb{Z}_{+}\text{.} \label{EqCompute24}%
\end{equation}

Specifically, $q_{0}=1$, $p_{1}=1$, $q_{1}=1+\xi$; and
\begin{align}
p_{n+1}  &  =p_{n}+q_{n}\label{EqCompute25}\\
q_{n+1}  &  =q_{n}+\xi^{n+1}p_{n+1}=\xi^{n+1}p_{n}+\left(  1+\xi^{n+1}\right)
q_{n}\text{.} \label{EqCompute26}%
\end{align}
In matrix form:
\begin{multline}
\binom{p_{n}\left(  \xi\right)  }{q_{n}\left(  \xi\right)  }=\left(
\begin{array}
[c]{ll}%
1 & 1\\
\xi^{n} & 1+\xi^{n}%
\end{array}
\right)  \left(
\begin{array}
[c]{ll}%
1 & 1\\
\xi^{n-1} & 1+\xi^{n-1}%
\end{array}
\right) \label{EqCompute27}\\
\cdots\left(
\begin{array}
[c]{ll}%
1 & 1\\
\xi^{2} & 1+\xi^{2}%
\end{array}
\right)  \left(
\begin{array}
[c]{ll}%
1 & 1\\
\xi & 1+\xi
\end{array}
\right)  \left(
\begin{array}
[c]{l}%
0\\
1
\end{array}
\right)  \text{.}%
\end{multline}

The polynomials may well be of independent interest, and we will need the
first few in the infinite string:%

\begin{table}[h] \centering
$%
\begin{tabular}
[c]{lll}\hline
\multicolumn{1}{|l}{$n$} & \multicolumn{1}{|l}{$p_{n}\left(  \xi\right)  $} &
\multicolumn{1}{|l|}{$q_{n}\left(  \xi\right)  $}\\\hline
\multicolumn{1}{|l}{$n=1$} & \multicolumn{1}{|l}{$1$} &
\multicolumn{1}{|l|}{$1+\xi$}\\\hline
\multicolumn{1}{|l}{$n=2$} & \multicolumn{1}{|l}{$2+\xi$} &
\multicolumn{1}{|l|}{$%
\begin{array}
[c]{l}%
\xi^{2}+\left(  1+\xi^{2}\right)  \left(  1+\xi\right) \\
=1+\xi+2\xi^{2}+\xi^{3}%
\end{array}
$}\\\hline
\multicolumn{1}{|l}{$n=3$} & \multicolumn{1}{|l}{$3+2\xi+2\xi^{2}+\xi^{3}$} &
\multicolumn{1}{|l|}{$%
\begin{array}
[c]{l}%
\xi^{3}\left(  2+\xi\right)  +\left(  1+\xi^{3}\right)  \left(  1+\xi+2\xi
^{2}+\xi^{3}\right) \\
=1+\xi+2\xi^{2}+4\xi^{3}+2\xi^{4}+2\xi^{5}+\xi^{6}%
\end{array}
$}\\\hline
&  &
\end{tabular}
$\caption{~}\label{TableKey copy(3)}%
\end{table}%

\begin{lemma}
\label{LemCompute0}The first and the last terms in the polynomials
$p_{n}\left(  \xi\right)  $ and $q_{n}\left(  \xi\right)  $ are as indicated
in the following formulas\emph{:}
\begin{equation}
p_{n}\left(  \xi\right)  =n+\left(  n-1\right)  \xi+\cdots+\xi^{\frac{\left(
n-1\right)  n}{2}}\text{,} \label{EqComputeInsert1}%
\end{equation}
and
\begin{equation}
q_{n}\left(  \xi\right)  =1+\xi+\cdots+\xi^{\frac{n\left(  n+1\right)  }{2}%
}\text{.} \label{EqComputeInsert2}%
\end{equation}
So the degree of $q_{n}\left(  \xi\right)  $ is $\frac{n\left(  n+1\right)
}{2}$.
\end{lemma}

\begin{proof}
Note that Table 4 already suggests the start of an induction proof. Now
suppose (\ref{EqComputeInsert1}) and (\ref{EqComputeInsert2}) hold up to $n$,
i.e., $p_{n}\left(  0\right)  =n$; and that both $p_{n}\left(  \xi\right)  $
and $q_{n}\left(  \xi\right)  $ have leading coefficients one.

With the use of (\ref{EqCompute25}) and (\ref{EqCompute26}) we then get
\begin{align*}
p_{n+1}\left(  0\right)   &  =p_{n}\left(  0\right)  +q_{n}\left(  0\right) \\
&  =n+1\text{, }%
\end{align*}
which is the next step in the induction. Now apply the same argument to
\[
\left.  \frac{d}{d\xi}p_{n+1}\right\vert _{\xi=0}\text{,}%
\]
and the result in (\ref{EqComputeInsert1}) for the next coefficient follows.

Setting $\xi=0$ in
\begin{equation}
q_{n+1}\left(  \xi\right)  =\xi^{n+1}p_{n}\left(  \xi\right)  +\left(
1+\xi^{n+1}\right)  q_{n}\left(  \xi\right)  \label{EqComputeInsert3}%
\end{equation}
and using the induction hypothesis, we get $q_{n+1}\left(  0\right)
=q_{n}\left(  0\right)  =1$.

We now turn to the leading coefficient in $q_{n+1}\left(  \xi\right)  $. As
before, we use the induction hypothesis, and (\ref{EqComputeInsert3}). We get
\begin{align*}
q_{n+1}\left(  \xi\right)   &  =\xi^{n+1}\left(  n+\cdots+\xi^{\frac{\left(
n-1\right)  n}{2}}\right)  +\left(  1+\xi^{n+1}\right)  \left(  1+\cdots
+\xi^{\frac{n\left(  n+1\right)  }{2}}\right) \\
&  =1+\cdots+\xi^{\frac{\left(  n+1\right)  \left(  n+2\right)  }{2}}\text{.}%
\end{align*}
This completes the induction proof.

In other words, the degree of $q_{n}\left(  \xi\right)  $ is
\[
1+2+3+\cdots+n=\frac{n\left(  n+1\right)  }{2};
\]
and each $q_{n}\left(  \xi\right)  $ has leading coefficient one.
\end{proof}

\begin{proposition}
\label{PropCompute0}~\newline\noindent\emph{(}a\emph{)} The two generating
functions
\begin{equation}
P\left(  X,\xi\right)  \text{\emph{:}}=\sum_{n=1}^{\infty}p_{n}\left(
\xi\right)  X^{n}\text{,} \label{EqComputeInsert4}%
\end{equation}
and
\begin{equation}
Q\left(  X,\xi\right)  \text{\emph{:}}=\sum_{n=0}^{\infty}q_{n}\left(
\xi\right)  X^{n} \label{EqComputeInsert5}%
\end{equation}
satisfy
\begin{equation}
X\left(  P\left(  X,\xi\right)  +Q\left(  X,\xi\right)  \right)  =P\left(
X,\xi\right)  \label{EqComputeInsert6}%
\end{equation}
and
\begin{equation}
Q\left(  X,\xi\right)  =1+XQ\left(  X,\xi\right)  +P\left(  \xi X,\xi\right)
\text{.} \label{EqComputeInsert7}%
\end{equation}

\noindent\emph{(}b\emph{)} Here we have picked $\xi\in\left(  0,1\right)
$\emph{;} and we note that, in the first variable, $Q\left(  X,\xi\right)  $
has \emph{radius of convergence} $1$, while $P\left(  X,\xi\right)  $ has
radius of convergence $\sqrt{\xi}$.
\end{proposition}

\begin{proof}
Multiplying through by $X^{n}$ in
\[
p_{n+1}\left(  \xi\right)  =p_{n}\left(  \xi\right)  +q_{n}\left(  \xi\right)
\text{,}%
\]
and using
\begin{equation}
\binom{p_{0}}{q_{0}}=\binom{0}{1}\text{,} \label{EqComputeInsert8}%
\end{equation}
we arrive at the first formula (\ref{EqComputeInsert6}) in the statement of
the proposition.

For the proof of (\ref{EqComputeInsert7}) we again multiply through by
$X^{n+1}$, now in
\[
q_{n+1}\left(  \xi\right)  =q_{n}\left(  \xi\right)  +\xi^{n+1}p_{n+1}\left(
\xi\right)  \text{.}%
\]
After adding up the terms with \textquotedblleft$\sum_{n=0}^{\infty}\cdots
$\textquotedblright\ and using (\ref{EqComputeInsert8}), we arrive at the
desired conclusion (\ref{EqComputeInsert7}).

We now turn to the radii of convergence:

Since we already established that
\[
\sum_{n=1}^{\infty}\xi^{n}p_{n}\left(  \xi\right)  ^{2}<\infty\text{,}%
\]
it follows that there is a finite constant $C$ such that
\begin{equation}
p_{n}\left(  \xi\right)  \leq C\xi^{-\frac{n}{2}}\text{;}
\label{EqComputeInsert9}%
\end{equation}
and we conclude that $X\longmapsto P\left(  X,\xi\right)  $ has radius of
convergence $\sqrt{\xi}$ as claimed.

But further note that $n\longmapsto q_{n}\left(  \xi\right)  $ is bounded, as
a consequence of the estimate (\ref{EqComputeInsert9}). Hence we conclude that
$X\longmapsto Q\left(  X,\xi\right)  $ has radius of convergence $=1$.
\end{proof}

The purpose of the previous discussion is to find the deficiency vector $u$,
i.e., the solution $u$ in $\mathcal{H}_{E}$ in a random walk model with
$p\left(  n,n+1\right)  =\frac{M}{1+M}$, $p\left(  n,n-1\right)  =\frac
{1}{1+M}$ where $M>1$. We now turn to the corresponding generating functions
\[
Q\left(  X,\frac{1}{M}\right)  =\sum_{n=1}^{\infty}u\left(  n\right)
X^{n}\text{.}%
\]

\begin{corollary}
\label{CorCompute0}The \emph{generating function} $X\longmapsto P\left(
X,\xi\right)  $ from \emph{(}\ref{EqComputeInsert4}\emph{)} in Proposition
\ref{PropCompute0} has the representation
\begin{equation}
P\left(  X,\xi\right)  =\sum_{n=1}^{\infty}\frac{\xi^{\frac{n\left(
n+1\right)  }{2}}X^{n}}{\left(  1-X\right)  ^{2}\left(  1-\xi X\right)
^{2}\left(  1-\xi^{2}X\right)  ^{2}\cdots\left(  1-\xi^{n}X\right)  ^{2}%
}\text{.} \label{EqComputeInsert10}%
\end{equation}

\end{corollary}

\begin{proof}
This is an application of (\ref{EqComputeInsert4}) and (\ref{EqComputeInsert5}%
): Eliminate $Q\left(  X,\xi\right)  $ and iterate the substitution.
\end{proof}

\begin{corollary}
\label{CorCompute0.5}The generating function for $u$ itself is as
follows\emph{:}
\begin{align*}
Q\left(  X,\xi\right)   &  =\sum_{n=0}^{\infty}u\left(  n,\xi\right)  X^{n}\\
&  =\frac{1}{1-X}\left(  1+\sum_{n=1}^{\infty}\frac{\xi^{\frac{n\left(
n+1\right)  }{2}}X^{n}}{\left(  1-\xi X\right)  ^{2}\cdots\left(  1-\xi
^{n}X\right)  ^{2}}\right)  \text{,}%
\end{align*}
where
\[
\prod\limits_{k=1}^{\infty}\left(  1-\xi^{k}X\right)  =\exp\left(  -\frac{\xi
X}{1-X}\right)  \text{.}%
\]

\end{corollary}

\begin{proof}
Combine the previous formulas.
\end{proof}

In the next section, we take up a number of dynamics related issues concerning
these polynomials. Below we are concerned with the proof of the following:

\begin{lemma}
\label{LemCompute1}For every $\xi\in\left(  0,1\right)  $, there is an
$m\in\mathbb{N}$ such that
\begin{equation}
p_{x}\left(  \xi\right)  \leq x^{m}\text{,} \label{EqCompute28}%
\end{equation}
and
\begin{equation}
q_{x}\left(  \xi\right)  \leq\left(  x+1\right)  ^{m}-x^{m}\text{,~for all
}x\in\mathbb{Z}_{+}\text{,} \label{EqCompute29}%
\end{equation}
where $m=m\left(  \xi\right)  $ depends on $\xi$.
\end{lemma}

\begin{proof}
Table 4 makes clear the start of an induction of (\ref{EqCompute28}) and
(\ref{EqCompute29}). Now suppose $m$ has been chosen such that
(\ref{EqCompute28}) and (\ref{EqCompute29}) hold up to $x$.

Then
\begin{align*}
p_{x+1}\left(  \xi\right)   &  =p_{x}\left(  \xi\right)  +q_{x}\left(
\xi\right) \\
&  \leq x^{m}+\left(  x+1\right)  ^{m}-x^{m}\\
&  =\left(  x+1\right)  ^{m}\text{.}%
\end{align*}
As a result, we get
\begin{align}
q_{x+1}\left(  \xi\right)   &  =q_{x}\left(  \xi\right)  +\xi^{x+1}%
p_{x+1}\left(  \xi\right) \label{EqCompute30}\\
&  \leq\left(  x+1\right)  ^{m}-x^{m}+\xi^{x+1}\left(  x+1\right)
^{m}\text{.}\nonumber
\end{align}

Now in the next step, we adjust $m$ such that
\[
\left(  x+1\right)  ^{m}-x^{m}+\xi^{x+1}\left(  x+1\right)  ^{m}\leq\left(
x+2\right)  ^{m}-\left(  x+1\right)  ^{m}\text{.}%
\]

We rewrite this:
\begin{align*}
2+\xi^{x+1}  &  \leq\left(  1-\frac{1}{x+1}\right)  ^{m}+\left(  1+\frac
{1}{x+1}\right)  ^{m}\\
&  \leq2\cdot\left(  1+\binom{m}{2}\left(  \frac{1}{x+1}\right)  ^{2}\right)
\text{;}%
\end{align*}
or
\begin{equation}
\left(  x+1\right)  ^{2}\xi^{x+1}\leq m\left(  m-1\right)  \text{.}
\label{EqCompute31}%
\end{equation}
But
\begin{equation}
\max\limits_{t\in\mathbb{R}_{+}}t^{2}\xi^{t}=\left(  \frac{2}{\ln\xi}\right)
^{2}e^{-2}\text{.} \label{EqCompute32}%
\end{equation}

It follows that (\ref{EqCompute31}) holds if
\[
\left\vert \ln\xi\right\vert >\frac{2}{e\sqrt{m\left(  m-1\right)  }}\text{,}%
\]
and so
\begin{equation}
M>\exp\left(  \frac{2}{e\sqrt{m\left(  m-1\right)  }}\right)  \text{.}
\label{EqCompute33}%
\end{equation}

Since the limit on the RHS in (\ref{EqCompute33}) is $1$ as $m\rightarrow
\infty$, and $M>1$ is fixed, it follows that $m$ can be adjusted to $\xi=1/M$
such that (\ref{EqCompute33}) holds. With this choice in fact
(\ref{EqCompute29}) will be satisfied for all $x\in\mathbb{Z}_{+}$. To see
this, apply (\ref{EqCompute32}) to $t=x+1$.
\end{proof}

\begin{lemma}
\label{LemCompute2}The solution $u$ from Part 3 in Table 3 satisfies
\begin{equation}
\mathcal{E}\left(  u\right)  <\infty\label{EqCompute34}%
\end{equation}

\end{lemma}

\begin{proof}%
\begin{align*}
\mathcal{E}\left(  u\right)   &  =\sum_{x\in\mathbb{Z}_{+}}\mu\left(
x\right)  \left(  \left(  \delta u\right)  \left(  x\right)  \right)  ^{2}\\
&  =_{\left(  \text{by Lemma \ref{LemCompute2}}\right)  }\sum_{x\in
\mathbb{Z}_{+}}M^{x}\left(  \xi^{x}p_{x}\left(  \xi\right)  \right)  ^{2}\\
&  \leq\sum_{x\in\mathbb{Z}_{+}}x^{2m}\xi^{x}<\infty\text{.}%
\end{align*}

\end{proof}

The finishes the proof of Part Three in Theorem \ref{TheoCompute1}.\medskip

\textbf{Part 4.} Here the model is $\left(  \mathbb{Z},\mu\right)  $ when the
function $\mu$ satisfies (\ref{EqCompute6}) and (\ref{EqCompute7}). As a
result the solution $u$ to $\Delta u=-u$ considered in the theorem satisfies
\begin{equation}
u\left(  x\right)  =u\left(  -x\right)  ,~x\in\mathbb{Z}\text{.}
\label{EqCompute35}%
\end{equation}
We get
\[
\left(  \Delta u\right)  \left(  0\right)  =M\left(  2u\left(  0\right)
-u\left(  1\right)  \right)  =-2M\left(  \delta u\right)  \left(  1\right)
=-u\left(  0\right)  \text{.}%
\]
Hence
\begin{equation}
\left(  \delta u\right)  \left(  1\right)  =\left(  \frac{\xi}{2}\right)
u\left(  0\right)  \text{,} \label{EqCompute36}%
\end{equation}
and
\[
u\left(  1\right)  =\left(  1+\frac{\xi}{2}\right)  u\left(  0\right)
\text{.}%
\]
Now $u\left(  x\right)  $ is the second component in the vector
\begin{multline*}
\binom{\ast}{u\left(  x\right)  }=\left(
\begin{array}
[c]{ll}%
1 & 1\\
\xi^{x} & 1+\xi^{x}%
\end{array}
\right)  \left(
\begin{array}
[c]{ll}%
1 & 1\\
\xi^{x-1} & 1+\xi^{x-1}%
\end{array}
\right) \\
\cdots\left(
\begin{array}
[c]{ll}%
1 & 1\\
\xi^{2} & 1+\xi^{2}%
\end{array}
\right)  \binom{\frac{1}{2}}{1+\frac{\xi}{2}}u\left(  0\right)  \text{.}%
\end{multline*}
\qed

\section{The Polynomials $p_{n}\left(  z\right)  $ and $q_{n}\left(  z\right)
,$ $n=1,2,\cdots$\label{Poly}}

We begin with some technical lemmas, and we further observe that the examples
from section \ref{Compute} indeed have a more general flavor: for example
(Proposition \ref{PropRandom1}), they may be derived from a standard random
walk model. We will do the computations here just for a one-dimensional walk,
but the basic idea carries over much more generally as we show in the next section.

\textsc{Optimal Estimates.} In the previous section, we considered $M>1,$
$\xi=1/M,$ and two sequences $p_{n}\left(  \xi\right)  ,$ $q_{n}\left(
\xi\right)  $ for $n=1,2,3,\cdots$. In the proof of Theorem \ref{TheoCompute1}%
, we showed the following:

\begin{proposition}
\label{PropPoly1}If
\begin{equation}
M>\exp\left(  e^{-1}\sqrt{2/3}\right)  , \label{EqPoly1}%
\end{equation}
then
\begin{equation}
p_{n}\left(  \xi\right)  \leq n^{3}\text{ for all }n\in\mathbb{Z}_{+}\text{.}
\label{EqPoly2}%
\end{equation}

\end{proposition}

In the following sense, this is not optimal; in fact, there is a finite
constant $B$ such that we get linear bounds in both directions. Specifically:
\begin{equation}
n\leq p_{n}\left(  \xi\right)  \leq1+B\left(  \frac{\sqrt{\xi}}{1-\sqrt{\xi}%
}\right)  n\text{.} \label{EqPoly3}%
\end{equation}

\begin{proof}
We will establish the existence of $B\left(  \in\mathbb{R}_{+}\right)  $ by
induction; and the size of $B$ will follow from the \textit{a priori}
estimates to follow. The induction begins with an inspection of Table 4 in
section \ref{Compute}.

\begin{lemma}
\label{LemPoly1}For every $\xi\in\left(  0,1\right)  $, the following finite
limit exists\emph{:}
\begin{equation}
Q\left(  \infty,\xi\right)  \text{\emph{:}}=\lim\limits_{n\rightarrow\infty
}q_{n}\left(  \xi\right)  ; \label{EqPoly4}%
\end{equation}
the limit is monotone, and $Q\left(  \infty,\xi\right)  >1$.

\begin{proof}
Let $u=u_{\xi}$ be the function from Lemma \ref{LemCompute1}. \newline

We established that
\[
\mathcal{E}\left(  u_{\xi}\right)  =\sum_{n=1}^{\infty}\xi^{n}p_{n}\left(
\xi\right)  ^{2}<\infty
\]
where $\left(  \delta u_{\xi}\right)  \left(  n\right)  =\xi^{n}p_{n}\left(
\xi\right)  $. If $\xi^{n}p_{n}\left(  \xi\right)  ^{2}\leq A,~\forall
n\in\mathbb{Z}_{+}$, for some fixed constant $A$\emph{;} then
\begin{equation}
\left(  \delta u_{\xi}\right)  \left(  n\right)  \leq\sqrt{A}\xi^{n/2}\text{.}
\label{EqPoly5}%
\end{equation}
Since
\begin{equation}
u_{\xi}\left(  n\right)  =1+\left(  \delta u_{\xi}\right)  \left(  1\right)
+\cdots+\left(  \delta u_{\xi}\right)  \left(  n\right)  \text{,}
\label{EqPoly6}%
\end{equation}
we get
\begin{align*}
u_{\xi}\left(  n\right)   &  \leq1+\sqrt{A\xi}\frac{1-\xi^{n/2}}{1-\sqrt{\xi}%
}\\
&  <1+\frac{\sqrt{A\xi}}{1-\sqrt{\xi}}\text{.}%
\end{align*}
Since
\begin{equation}
u_{\xi}\left(  1\right)  <u_{\xi}\left(  2\right)  <\cdots<u_{\xi}\left(
n\right)  <u_{\xi}\left(  n+1\right)  \cdots\text{,} \label{EqPoly7}%
\end{equation}
the desired conclusion follows, and
\begin{equation}
Q\left(  \infty,\xi\right)  \leq1+\frac{\sqrt{A\xi}}{1-\sqrt{\xi}}\text{.}
\label{EqPoly8}%
\end{equation}
We will see below that
\[
p_{n}\left(  \xi\right)  =1+\sum_{x=1}^{n-1}q_{x}\left(  \xi\right)  \text{,}%
\]
so
\[
p_{n}\left(  \xi\right)  \leq1+\left(  \frac{\sqrt{A\xi}}{1-\sqrt{\xi}%
}\right)  n\text{.}%
\]
The estimate in the RHS in (\ref{EqPoly6}) now follows. \newline

Returning to $p_{n}\left(  \xi\right)  $, we have
\begin{align}
p_{n}\left(  \xi\right)   &  =p_{n-1}\left(  \xi\right)  +q_{n-1}\left(
\xi\right) \label{EqPoly9}\\
&  =q_{n-1}\left(  \xi\right)  +q_{n-2}\left(  \xi\right)  +\cdots
+q_{1}\left(  \xi\right)  +p_{1}\left(  \xi\right) \nonumber
\end{align}
where
\[
\binom{p_{1}\left(  \xi\right)  }{q_{1}\left(  \xi\right)  }=\binom{1}{1+\xi
}.
\]
Since
\[
q_{x}\left(  \xi\right)  \geq1\text{ for }x\in\mathbb{Z}_{+}\text{,}%
\]
(\ref{EqPoly9}) implies
\[
p_{n}\left(  \xi\right)  \geq n\text{,}%
\]
which is the remaining lower boundary on the LHS in (\ref{EqPoly6}).
\end{proof}
\end{lemma}
\end{proof}

\section{A Random Walk Model\label{Random}}

In this section we resume our analysis of the general case: How does the
operator theory throw light on \textquotedblleft the boundary at
infinity\textquotedblright\ as it was made precise in section \ref{Compute}?
Part of the answer lies in a transfer operator studied in Lemma
\ref{LemRandom2} below.

For the weighted graphs $\left(  G,c\right)  $ in section \ref{Frame} with
$c:G^{1}\rightarrow\mathbb{R}_{+}$ representing conductance we may introduce
\[
c\left(  x\right)  =\sum_{y\sim x}c\left(  x,y\right)
\]
and
\begin{equation}
p\left(  x,y\right)  \text{:}=\frac{c\left(  x,y\right)  }{c\left(  x\right)
}\text{.} \label{EqRandom1}%
\end{equation}
Then
\begin{equation}
\sum_{y\sim x}p\left(  x,y\right)  =1 \label{EqRandom2}%
\end{equation}
and the function $p\left(  x,y\right)  $ represents a system of transition
probabilities. They in turn determine a random walk on the set $G^{0}$ of all
vertices. If $\omega$ is a path of a \textquotedblleft
walker\textquotedblright, and $X_{n}\left(  \omega\right)  $ is the location
in $G^{0}$ at time $n$, then
\begin{equation}
\operatorname*{Prob}\left(  \left\{  \omega|X_{n}\left(  \omega\right)
=x,X_{n+1}\left(  \omega\right)  =y\right\}  \right)  =p\left(  x,y\right)
\text{ for all }\left(  x,y\right)  \in G^{1}\text{,} \label{EqRandom3}%
\end{equation}
where \textquotedblleft Prob\textquotedblright\ is short for \textquotedblleft
probability.\textquotedblright\ This is a \textit{reversible} random walk in
that
\begin{equation}
c\left(  x\right)  p\left(  x,y\right)  =c\left(  y\right)  p\left(
y,x\right)  \text{ for all }x\sim y\text{.} \label{EqRandom4}%
\end{equation}

\begin{proposition}
\label{PropRandom1}In the model $\left(  \mathbb{Z}_{+},M^{x}\right)  $ in
Theorem \ref{TheoCompute1}, the transition probability from $0$ to $1$ is $1$,
and from $n$ to $n+1$, it is the fixed quotient $^{M}/_{1+M}$ for all
$n\in\mathbb{Z}_{+}$. Specifically, a move to the right\emph{:}
\[
p\left(  n,n+1\right)  =\frac{M}{1+M};
\]
and
\[
p\left(  n,n-1\right)  =\frac{1}{1+M}%
\]
equals the probability of a move to the left when starting at the vertex $n$.
\end{proposition}

\begin{proof}
Since the edges in $\mathbb{Z}_{+}\cup\left\{  0\right\}  $ are just the
nearest neighbors, we get
\begin{align*}
p\left(  0,1\right)   &  =1\text{, and}\\
p\left(  n,n+1\right)   &  =\frac{M^{n+1}}{M^{n}+M^{n+1}}\\
&  =\frac{M^{n+1}}{M^{n}\cdot\left(  1+M\right)  }\\
&  =\frac{M}{1+M}\text{.}%
\end{align*}

\end{proof}

\begin{proposition}
\label{PropRandom2}In the model on $\mathbb{Z}$ with the following
\[
c\left(  0,1\right)  =A,c\left(  n-1,n\right)  =A^{n},~n\in\mathbb{Z}_{+};
\]
and
\[
c\left(  -1,0\right)  =B
\]
and
\[
c\left(  -n,-n+1\right)  =B^{n}%
\]
we get the following transition probabilities\emph{:}
\begin{align*}
p\left(  0,1\right)   &  =\frac{A}{A+B},\\
p\left(  0,-1\right)   &  =\frac{B}{A+B},\\
p\left(  n,n+1\right)   &  =\frac{A}{1+A},\\
p\left(  n,n-1\right)   &  =\frac{1}{1+A},~\text{for }n\in\mathbb{Z}%
_{+};\text{ and}\\
p\left(  -n,-n-1\right)   &  =\frac{B}{1+B},\text{ and }\\
p\left(  -n,-n+1\right)   &  =\frac{1}{1+B}\text{.}%
\end{align*}

\end{proposition}

\begin{proof}
Left to the reader.
\end{proof}

The significance of the deficiency vectors $u$ studied in section \ref{Frame}
is that the presence of a non-zero deficit eigenspace
\begin{equation}
\operatorname*{Def}\text{:}=\left\{  u\in\mathcal{H}_{E}\,|\,\Delta
u=-u\right\}  \label{EqRandom5}%
\end{equation}
offers a quantitative measure of states at \textquotedblleft
infinity\textquotedblright\ in the model.

To make this precise, we return to the operator theory for the operator
$\Delta$ with its dense domain $\mathcal{D}$ in $\mathcal{H}_{E}$; see
Proposition \ref{PropFrame1} for details.

\begin{definition}
\label{DefRandom1}The domain of the adjoint operator $\Delta^{\ast}$ is
defined as follows
\begin{equation}
\operatorname*{dom}\left(  \Delta^{\ast}\right)  =\left\{  u\in\mathcal{H}%
_{E}|\text{ s.t. }\exists C>\infty\text{ with }\left\vert \left\langle
u,\Delta\varphi\right\rangle \right\vert \leq C\left\Vert \varphi\right\Vert
,~\forall\varphi\in\mathcal{D}\text{.}\right\}  \label{EqRandom6}%
\end{equation}
If $u\in\operatorname*{dom}\left(  \Delta^{\ast}\right)  $, set $\Delta^{\ast
}u=v$ if
\begin{equation}
\left\langle v,\varphi\right\rangle =\left\langle u,\Delta\varphi\right\rangle
\text{ holds for all }\varphi\in\mathcal{D}\text{.} \label{EqRandom7}%
\end{equation}

\end{definition}

We say that $\Delta$ is essentially selfadjoint if its closure $\Delta^{clo}$
is selfadjoint, i.e., if $\Delta^{clo}=\Delta^{\ast}$.

Now $\Delta$ is semibounded on its domain $\mathcal{D}$ in $\mathcal{H}_{E}$
by Proposition \ref{PropFrame1}; and it thus follows from a theorem by von
Neumann \cite{DuSc88} that $\Delta$ is essentially selfadjoint iff the
corresponding deficiency space $\operatorname*{Def}$ is zero. To make the
connection between the operator theory and the solutions $u$ studied in
section \ref{Compute}, we will need the following.

\begin{lemma}
\label{LemRandom1}Let $\mathcal{H}_{E},\Delta,\mathcal{D}$ and $\Delta^{\ast}$
be as in Definition \ref{DefRandom1}. If $u\in\operatorname*{dom}\left(
\Delta^{\ast}\right)  $, then
\begin{equation}
\left(  \Delta^{\ast}u\right)  \left(  x\right)  =\sum_{y\sim x}c\left(
x,y\right)  \left(  u\left(  x\right)  -u\left(  y\right)  \right)  \text{.}
\label{EqRandom8}%
\end{equation}

\end{lemma}

\begin{proof}
Note that the assumptions in the lemma include the following two assertions:

\begin{enumerate}
\item[(i)] $u\in\mathcal{H}_{E}$, and

\item[(ii)] $\Delta^{\ast}u\in\mathcal{H}_{E}$.
\end{enumerate}

Now let $y\in G^{0}\diagdown\left(  0\right)  $, $and$ compute
\begin{align*}
\left\langle v_{y},\Delta^{\ast}u\right\rangle _{E}  &  =\left\langle \Delta
v_{y},u\right\rangle _{E}\\
&  =\left\langle \delta_{y}-\delta_{o},u\right\rangle _{E}\\
&  =\sum_{z\sim y}c\left(  y,z\right)  \left(  u\left(  y\right)  -u\left(
z\right)  \right)  -\sum_{z\sim o}c\left(  o,z\right)  \left(  u\left(
o\right)  -u\left(  z\right)  \right) \\
&  =\left\langle v_{y}\left(  \cdot\right)  ,\sum_{z}c\left(  \cdot,z\right)
\left(  u\left(  \cdot\right)  -u\left(  z\right)  \right)  \right\rangle
_{E}\text{.}%
\end{align*}

Since the family $\left(  v_{y}\right)  $ spans the dense subspace
$\mathcal{D}$ in $\mathcal{H}_{E}$, the desired formula (\ref{EqRandom8}) now
follows. Note that (\ref{EqRandom8}) is the identity of two vectors in the
energy Hilbert space $\mathcal{H}_{E}$.
\end{proof}

For the two-sided random walk on $\mathbb{Z}\,$with conductances given in
Figure 1 below, we are now ready to compute the deficit space
$\operatorname*{Def}$ in $\mathcal{H}_{E}$.
\[%
{\parbox[b]{5.0286in}{\begin{center}
\includegraphics[
natheight=0.947300in,
natwidth=2.880000in,
height=1.6845in,
width=5.0286in
]%
{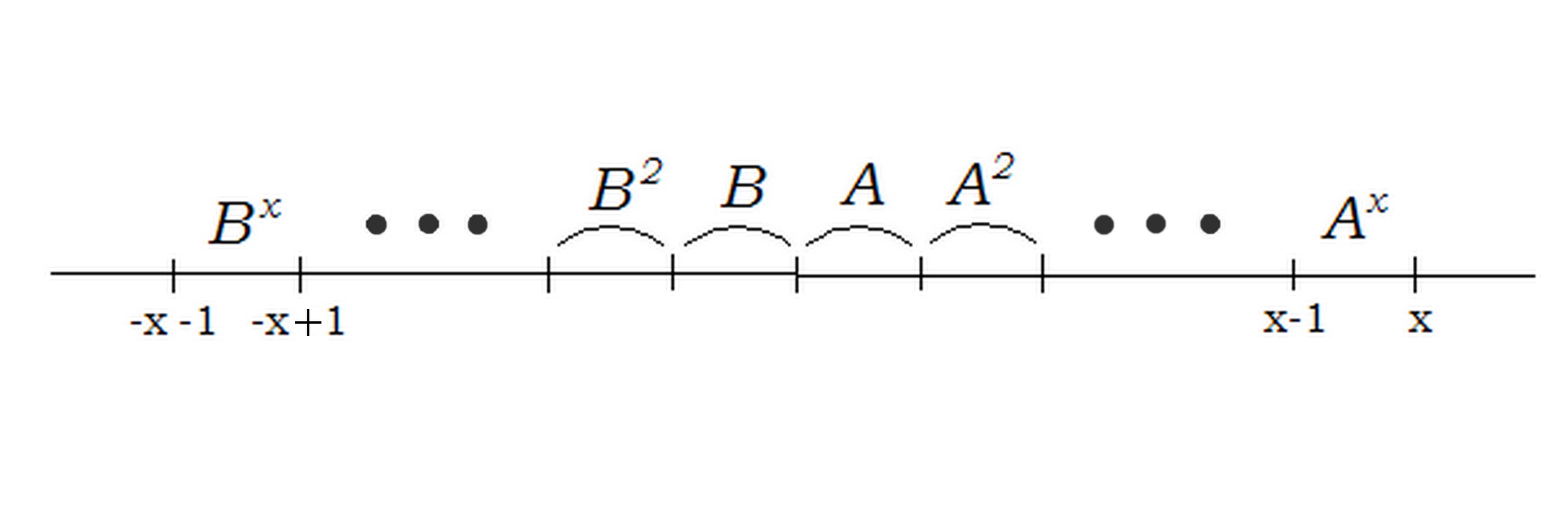}%
\\
\bfseries{Figure 1.} Conductance.
\end{center}}}%
\]

\begin{proposition}
\label{PropRandom3}The deficit vectors $u\in\mathcal{H}_{E}$ solving
$\Delta^{\ast}u=-u$ in the $A-B$ model on $\mathbb{Z}\,$given in Figure 1 are
determined by the following system\emph{:}
\begin{equation}
\left(  A+B+1\right)  u\left(  0\right)  =Au\left(  1\right)  +Bu\left(
-1\right)  \label{EqRandom9}%
\end{equation}
and
\begin{equation}
p_{1}\left(  \alpha\right)  +p_{1}\left(  \beta\right)  =1\text{,}
\label{EqRandom10}%
\end{equation}
where $\alpha=\,^{1}/_{A}$, and $\beta=\,^{1}/_{B}$.

Setting $\mu\left(  0\right)  =1$, we then obtain the following recursive
determination: If $n\in\mathbb{Z}_{+}$, then
\[
\mu\left(  n\right)  =q_{n}\left(  \alpha\right)  ,u\left(  -n\right)
=q_{n}\left(  \beta\right)
\]
with
\begin{align}
\left(
\begin{array}
[c]{l}%
p_{n}\left(  \xi\right) \\
q_{n}\left(  \xi\right)
\end{array}
\right)   &  =\left(
\begin{array}
[c]{ll}%
1 & 1\\
\xi^{n} & 1+\xi^{n}%
\end{array}
\right)  \cdots\left(
\begin{array}
[c]{ll}%
1 & 1\\
\xi^{2} & 1+\xi^{2}%
\end{array}
\right)  \left(
\begin{array}
[c]{ll}%
1 & 1\\
\xi & 1+\xi
\end{array}
\right)  \left(
\begin{array}
[c]{l}%
0\\
1
\end{array}
\right) \nonumber\\
&  \mathstrut\label{EqRandom11}%
\end{align}
for $\xi=\alpha$ on the $\mathbb{Z}_{+}$ side, and $\xi=\beta$ on the
$\mathbb{Z}_{-}$ side.
\end{proposition}

\begin{proof}
By the lemma, we get the following identities:
\[
A\left(  \delta u\right)  \left(  1\right)  +B\left(  \delta u\right)  \left(
-1\right)  =u\left(  0\right)  \text{.}%
\]
With $u\left(  0\right)  =1$, this amounts to equations (\ref{EqRandom9}) and
(\ref{EqRandom10}).

Using
\[
u\left(  n\right)  =u\left(  n-1\right)  +\left(  \delta u\right)  \left(
n\right)  ,~n\in\mathbb{Z}_{+}\text{,}%
\]
we get
\[
u\left(  n\right)  =q_{n}\left(  \alpha\right)
\]
with
\[
q_{n}\left(  \alpha\right)  =q_{n-1}\left(  \alpha\right)  +\alpha^{n}%
p_{n}\left(  \alpha\right)  \text{.}%
\]
With
\[
u\left(  -n\right)  =u\left(  -n+1\right)  +\left(  \delta u\right)  \left(
-n\right)  \text{,}%
\]
we get $\left(  n\in\mathbb{Z}_{+}\right)  $:
\[
u\left(  -n\right)  =q_{n}\left(  \beta\right)
\]
with
\[
q_{n}\left(  \beta\right)  =q_{n-1}\left(  \beta\right)  +\beta^{n}%
p_{n}\left(  \beta\right)  \text{.}%
\]

The resulting formulas (\ref{EqRandom11}) (in matrix form) now follow from
this; and as a result the solution $u$ is determined on all of $\mathbb{Z}$,
with the normalizations $u\left(  0\right)  =1$, and (\ref{EqRandom10}).
\end{proof}

\begin{lemma}
\label{LemRandom2}Let $\left(  G,c\right)  $ be a weighted graph with
probabilities
\begin{equation}
p\left(  x,y\right)  \text{\emph{:}}=\frac{c\left(  x,y\right)  }{c\left(
x\right)  }, \label{EqRandom12}%
\end{equation}
and set
\begin{equation}
\left(  Tf\right)  \left(  x\right)  =\sum_{y}p\left(  x,y\right)  f\left(
y\right)  ; \label{EqRandom13}%
\end{equation}
$T$ for transfer operator.

\begin{enumerate}
\item[(a)] Then
\begin{equation}
Tf=f-\frac{1}{c}\Delta f\text{.} \label{EqRandom14}%
\end{equation}
A vector $f\in\mathcal{H}_{E}$ is in $\operatorname*{Def}$ if and only if
\begin{equation}
Tf=\left(  1+\frac{1}{c}\right)  f \label{EqRandom15}%
\end{equation}
where
\[
c\left(  x\right)  =\sum_{y\sim x}c\left(  x,y\right)  \text{.}%
\]

\item[(b)] In the example of Proposition \ref{PropRandom1} $(G^{0}=\left\{
0\right\}  \cup\mathbb{Z}_{+})$,
\begin{equation}
\left(  Tf\right)  \left(  x\right)  =\frac{1}{1+M}f\left(  x-1\right)
+\frac{M}{1+M}f\left(  x+1\right)  ,~\text{for all }x\in G^{0}\text{.}
\label{EqRandom 16}%
\end{equation}

\item[(c)] In general if $f\in\operatorname*{Def}$ is bounded, then the
pointwise limit
\begin{equation}
\lim\limits_{k\rightarrow\infty}\left(  T^{k}f\right)  \left(  x\right)
=\emph{:~}h\left(  x\right)  \label{EqRandom17}%
\end{equation}
exists, and $h$ is harmonic.
\end{enumerate}
\end{lemma}

\begin{proof}

\begin{enumerate}
\item[(a)] We have
\begin{align*}
\left(  \Delta f\right)  \left(  x\right)   &  =\sum_{y}c\left(  x,y\right)
\left(  f\left(  x\right)  -f\left(  y\right)  \right) \\
&  =c\left(  x\right)  f\left(  x\right)  -\sum_{y}c\left(  x,y\right)
f\left(  y\right) \\
&  =c\left(  x\right)  \left(  f\left(  x\right)  -\left(  Tf\right)  \left(
x\right)  \right)  \text{,}%
\end{align*}
and (\ref{EqRandom14}) follows.

\item[(b)] follows from (a) and substitution into (\ref{EqRandom12}), i.e.,
\[
c\left(  n-1,n\right)  =M^{n},\text{ for }n\in\mathbb{Z}_{+}.
\]

\item[(c)] follows from (a) and (b).
\end{enumerate}

In particular if the $\Delta$ a constant $A\in\mathbb{R}_{+}$ such that
$f\left(  x\right)  \leq A$ for all $x\in G^{0}$, then apply $T$ to the
positive function $A-f$, and we get $A\geq T^{k}f$ pointwise on $G^{0}$. Since
$Tf=\left(  1+\frac{1}{c}\right)  f$ we get
\[
f\leq Tf\leq T^{2}f\leq\cdots\leq T^{k}f\leq\cdots\leq A\text{.}%
\]

Therefore the limit $h\left(  x\right)  $ in (\ref{EqRandom17}) exists if and
only if $f\left(  \in\operatorname*{Def}\right)  $ is a bounded function.

The following is then immediate:
\[
\Delta h=0\Leftrightarrow Th=h\text{.}%
\]

\end{proof}

\section{The space $\operatorname*{Def}$ and boundary conditions\label{Space}}

In this section, we discuss the distinction between the operator and $\Delta$
considered in the Hilbert space $\mathcal{H}_{E}$ as compared with $\ell^{2}%
$:$=\ell^{2}\left(  G^{0}\right)  $. As before $\left(  G,c\right)  $ is a
given weighted graph with $c:G^{1}\rightarrow\mathbb{R}_{+}$ representing
conductances on the edges; and $G^{0}$ density the set of vertices. The
corresponding Laplace operator is discussed in sections \ref{Frame} and
\ref{Compute}, and we recall that it depends on the choice of function $c$.

When $\Delta$ is viewed as an operator in $\mathcal{H}_{E}$, we take as its
domain
\begin{equation}
\mathcal{D}_{V}\text{:}=\operatorname*{span}\left\{  v_{x}|x\in G^{0}%
\,\diagdown\left\{  o\right\}  \right\}  \label{EqSpace1}%
\end{equation}
where $o$ is fixed in $G^{0}$.

When $\Delta$ is viewed as an operator in $\ell^{2}$, we take as its domain
\begin{equation}
\mathcal{D}_{E}\text{:}=\text{ all finitely supported functions on }%
G^{0}\text{.} \label{EqSpace2}%
\end{equation}

To stress the distinction between the two cases $\mathcal{H}_{E}$ and
$\ell^{2}$, we shall use the following terminology, $\left(  \Delta
,\mathcal{D}_{V},\mathcal{H}_{E}\right)  $ and $\left(  \Delta,\mathcal{D}%
_{F},\ell^{2}\right)  $. The corresponding inner products will be
distinguished with subscripts $\left\langle \cdot,\cdot\right\rangle _{E}$ for
$\mathcal{H}_{E}$, and $\left\langle \cdot,\cdot\right\rangle _{2}$ or
$\left\langle \cdot,\cdot\right\rangle _{\ell^{2}}$ for the second, i.e.,
\begin{equation}
\left\langle u,v\right\rangle _{\ell^{2}}\text{:}=\sum_{x\in G^{0}}%
\overline{u\left(  x\right)  }\,v\left(  x\right)  \text{.} \label{EqSpace3}%
\end{equation}

We note that \textit{symmetry} and \textit{semiboundedness} is satisfied for
$\Delta$ with respect to the two inner products, i.e.,
\begin{equation}
\left\langle u,\Delta v\right\rangle _{E}=\left\langle \Delta u,v\right\rangle
_{E}\text{, and }\left\langle u,\Delta u\right\rangle _{E}\geq0\text{ for all
}u,v\in\mathcal{D}_{V}\text{;} \label{EqSpace4}%
\end{equation}
as well as
\begin{equation}
\left\langle u,\Delta v\right\rangle _{2}=\left\langle \Delta u,v\right\rangle
_{2}\text{, and }\left\langle u,\Delta u\right\rangle _{2}\geq0\text{, for all
}u,v\in\mathcal{D}_{F}\text{.} \label{EqSpace5}%
\end{equation}

\begin{lemma}
\label{LemSpace1}\cite{JoPe08} The operator $\left(  \Delta,\mathcal{D}%
_{F},\ell^{2}\right)  $ is essentially selfadjoint\emph{;} and so in
particular, there are no non-zero solutions to the deficit equations in
$\ell^{2}$\emph{:} If $\Delta u=-u$, and $u\in\ell^{2}$, then $u=0$.
\end{lemma}

By contrast, we saw in sections \ref{Compute} and \ref{Poly}, that
\begin{equation}
\Delta u=-u \label{EqSpace6}%
\end{equation}
has non-zero solutions in $\mathcal{H}_{E}$ for a rich class of unbounded
functions $c$ on $G^{1}$. And so in particular, $\left(  \Delta,\mathcal{D}%
_{V},\mathcal{H}_{E}\right)  $ is \textit{not} essentially selfadjoint.

\begin{proposition}
\label{PropSpace1}If $u\in\operatorname*{Def}\diagdown\left\{  0\right\}  $,
then the infinite sum
\begin{equation}
S_{2}\left(  u\right)  \text{\emph{:}}=\sum_{x\in G^{0}}\overline{u\left(
x\right)  }\left(  \Delta u\right)  \left(  x\right)  \label{EqSpace7}%
\end{equation}
is divergent.
\end{proposition}

\begin{proof}
Recall that $\Delta u=\Delta^{\ast}u=-u$ if $u$ is in $\operatorname*{Def}$
and ,as a result,
\[
S_{2}\left(  u\right)  =-\sum_{x}\left\vert u\left(  x\right)  \right\vert
^{2}\text{.}%
\]
But, by Lemma \ref{LemSpace1}, the solution $u$ to (\ref{EqSpace6}) cannot be
in $\ell^{2}$ and, as a result,
\[
\sum_{x}\left\vert u\left(  x\right)  \right\vert ^{2}=\infty\text{.}%
\]

\end{proof}

\begin{theorem}
\label{TheoSpace1}If $u\in\mathcal{H}_{E}\ominus\operatorname*{Def}$, then the
sum $S_{2}\left(  u\right)  $ in \emph{(}\ref{EqSpace7}\emph{)} is finite and
it has the following representation\emph{:}

There is a unique $v\in\operatorname*{dom}\left(  \Delta_{V}^{clo}\right)  $
such that
\[
u=v+\Delta_{V}^{clo}v\text{,}%
\]
and
\begin{equation}
\mathcal{E}\left(  u\right)  =S_{2}\left(  u\right)  +\mathcal{E}\left(
P_{\operatorname*{Harm}}v\right)  \text{,} \label{EqSpace8}%
\end{equation}
where $\mathcal{E}$ denotes the energy-form, i.e., $\mathcal{E}\left(
\cdot\right)  $\emph{:}$=\left\Vert \cdot\right\Vert _{\mathcal{H}_{E}}$ and
where $P_{\operatorname*{Harm}}$ denotes the projection onto the harmonic
functions $h$ in $\mathcal{H}_{E}$, i.e.,
\begin{equation}
\operatorname*{Harm}\text{\emph{:}}=\left\{  h\in\mathcal{H}_{E}|\Delta
h=0\right\}  \text{.} \label{EqSpace9}%
\end{equation}

\end{theorem}

\begin{proof}
We begin with the closed subspace $\operatorname*{Harm}$ in (\ref{EqSpace9}).

\begin{lemma}
\label{LemSpace2}\emph{(}\cite{JoPe08}\emph{)}
\begin{align}
\operatorname*{Harm}  &  =\left\{  h\in\mathcal{H}_{E}\,|\,h\bot\delta
_{x},~\forall x\in G^{0}\right\} \label{EqSpace10}\\
&  =\mathcal{H}_{E}\ominus\left\{  \delta_{x}\right\}  _{x\in G^{0}};\nonumber
\end{align}
or setting
\begin{equation}
\operatorname*{Fin}\text{\emph{:}}=\text{ the closed linear span of }\left(
\delta_{x}\right)  _{x\in G^{0}}\text{.} \label{EqSpace11}%
\end{equation}

\end{lemma}

\begin{proof}
We have for all $u\in\mathcal{H}_{E}$,
\begin{equation}
\left(  \Delta u\right)  \left(  x\right)  =\left\langle \delta_{x}%
,u\right\rangle _{E}\text{,} \label{EqSpace12}%
\end{equation}
and it follows that $\Delta u\equiv0$ iff $u$ satisfies the conditions in
(\ref{EqSpace10}).

If $P_{\operatorname*{Harm}}$ and $P_{\operatorname*{Fin}}$ denote the
respective projections onto the closed subspaces in (\ref{EqSpace10}) and
(\ref{EqSpace11}), then
\begin{equation}
P_{\operatorname*{Harm}}+P_{\operatorname*{Fin}}=I \label{EqSpace13}%
\end{equation}
where $I$ denotes the identity operator in $\mathcal{H}_{E}$.
\end{proof}

\begin{lemma}
\label{LemSpace3}Let $v,w\in\mathcal{D}_{v}$\emph{;} then
\begin{equation}
\sum_{x\in G^{0}}\overline{v\left(  x\right)  }\left(  \Delta w\right)
\left(  x\right)  =\left\langle P_{\operatorname*{Fin}}v,w\right\rangle
_{\mathcal{H}_{E}}\text{.} \label{EqSpace14}%
\end{equation}

\end{lemma}

\begin{proof}
Introduce the sum notation $\sum\bar{v}\Delta w$ for the left hand side
expression in (\ref{EqSpace14}). Then
\begin{align*}
\sum\overline{v\left(  x\right)  }\left(  \Delta w\right)  \left(  x\right)
&  =_{\left(  \text{by (\ref{EqSpace12})}\right)  }\sum_{x}\overline{v\left(
x\right)  }\left\langle \delta_{x},w\right\rangle _{E}\\
&  =\left\langle \sum_{x}v\left(  x\right)  \delta_{x},w\right\rangle _{E}\\
&  =_{\left(  \text{by (\ref{EqSpace11})}\right)  }\left\langle
P_{\operatorname*{Fin}}v,w\right\rangle _{E}\text{.}%
\end{align*}

\end{proof}

\begin{lemma}
\label{LemSpace4}If $v\in\operatorname*{dom}\left(  \Delta_{V}^{clo}\right)
$, then
\begin{equation}
\left\Vert v+\Delta v\right\Vert _{E}\geq\left\Vert v\right\Vert _{E}\text{,}
\label{EqSpace15}%
\end{equation}
and so
\begin{equation}
u\longmapsto\left(  I+\Delta\right)  ^{-1}u=v \label{EqSpace16}%
\end{equation}
is contractive on the closed subspace $\mathcal{H}_{E}\ominus
\operatorname*{Def}$ in $\mathcal{H}_{E}$.
\end{lemma}

\begin{proof}
We will be using the operator theoretic properties (\ref{EqSpace4}) and
(\ref{EqSpace5}) stated before Lemma \ref{LemSpace1}. In particular, we use
that if an operator $\Delta$ is hermitian symmetric and semibounded on a dense
domain in Hilbert space, the so is its closure $\Delta^{clo}=\Delta^{\ast\ast
}$. We will be using the same symbol $\Delta$ also when referring to the
closed operator.

Now for $v\in\operatorname*{dom}\left(  \Delta^{clo}\right)  $, we have
\begin{align*}
\left\Vert v+\Delta v\right\Vert _{E}^{2}  &  =\left\langle v+\Delta
v,v+\Delta v\right\rangle _{E}\\
&  =\left\Vert v\right\Vert _{E}^{2}+\left\langle \Delta v,v\right\rangle
_{E}+\left\langle v,\Delta v\right\rangle _{E}+\left\Vert \Delta v\right\Vert
_{E}^{2}\\
&  \geq\left\Vert v\right\Vert _{E}^{2}\text{ by (\ref{EqSpace4})}.
\end{align*}

\end{proof}

\noindent\textit{Proof of Theorem \ref{TheoSpace1} resumed.} We have
\begin{align}
\operatorname*{Def}  &  =N_{\mathcal{H}_{E}}\left(  I+\Delta_{v}^{\ast}\right)
\label{EqSpace17}\\
&  =\left(  R\left(  I+\Delta^{clo}\right)  \right)  ^{\bot}\text{.}\nonumber
\end{align}
By virtue of Lemma \ref{LemSpace4}, we further note that $R\left(
I+\Delta^{clo}\right)  $ is closed in $\mathcal{H}_{E}$, and as a result
\begin{equation}
R\left(  I+\Delta^{clo}\right)  =\mathcal{H}_{E}\ominus\operatorname*{Def}%
\text{.} \label{EqSpace18}%
\end{equation}
With $u=v+\Delta v$, we now compute as follows. For the expression $S_{2}$ in
equation (\ref{EqSpace7}) we have:
\begin{align*}
S_{2}\left(  u\right)   &  =\sum_{x}\bar{u}\Delta u\\
&  =\sum_{x}\bar{v}\Delta v+\sum_{x}\bar{v}\Delta^{2}v+\sum_{x}\overline
{\Delta v}\Delta v+\sum_{x}\overline{\Delta v}\Delta^{2}v\\
&  =_{\left(  \text{by Lemma (\ref{LemSpace3})}\right)  }\left\langle
P_{\operatorname*{Fin}}v,v\right\rangle _{E}+\left\langle
P_{\operatorname*{Fin}}v,\Delta v\right\rangle _{E}+\left\langle
P_{\operatorname*{Fin}}\Delta v,v\right\rangle _{E}+\left\Vert \Delta
v\right\Vert _{E}^{2}\\
&  =\left\Vert P_{\operatorname*{Fin}}v\right\Vert _{E}^{2}+\left\langle
v,\Delta v\right\rangle _{E}+\left\langle \Delta v,v\right\rangle
_{E}+\left\Vert \Delta v\right\Vert _{E}^{2}\\
&  =\left\Vert P_{\operatorname*{Fin}}v\right\Vert _{E}^{2}-\left\Vert
v\right\Vert _{E}^{2}+\left\Vert v+\Delta v\right\Vert _{E}^{2}\\
&  =_{\left(  \text{by (\ref{EqSpace12})}\right)  }-\left\Vert
P_{\operatorname*{Harm}}v\right\Vert _{E}^{2}+\left\Vert u\right\Vert _{E}%
^{2}\\
&  =-\mathcal{E}\left(  P_{\operatorname*{Harm}}v\right)  +\mathcal{E}\left(
u\right)  \text{.}%
\end{align*}
The desired conclusion (\ref{EqSpace8}) in the statement of the theorem now follows.
\end{proof}

\section{Embeddings of graphs and of Hilbert spaces\label{Embed}}

In this section we show that mappings between graphs (for example inclusions
of sub-graphs into an ambient super-graph) induce embeddings of the
corresponding Hilbert spaces. This is functorial, and it further allows
localization. By this we mean embeddings of certain local \textquotedblleft
portions\textquotedblright\ of an ambient infinite graph $G$ will produce
intertwining operators in such a way that local spectral information is
embedded into global. One advantage of this is that computations can more
easily be carried out in the \textquotedblleft local
portions\textquotedblright\ of $G$.

Following the definitions in section \ref{Frame}, we shall now consider pairs
of weighted graphs $\left(  G,c_{G}\right)  $ and $\left(  H,c_{H}\right)  $.
For the first one, we have a vertex set $G^{0}$ and corresponding edges
$G^{1}$; and similarly with $H^{0}$ and $H^{1}$ for the second graph. The
conductance functions $c_{G}$ and $c_{H}$ are as follows:
\begin{equation}
c_{G}:G^{1}\rightarrow\mathbb{R}_{+}\text{, and }c_{H}:H^{1}\rightarrow
\mathbb{R}_{+}\text{,} \label{EqEmbed1}%
\end{equation}
each satisfying the axioms from section \ref{Frame}. The corresponding energy
Hilbert spaces will be denoted $\mathcal{H}_{E}\left(  G,c_{G}\right)  $ and
$\mathcal{H}_{E}\left(  H,c_{H}\right)  $; or simply $\mathcal{H}_{E}\left(
G\right)  $ and $\mathcal{H}_{E}\left(  H\right)  $ when the choice of
conductance functions are clear from the context.

\begin{definition}
\label{DefEmbed1}Consider two functions
\begin{equation}
G^{0}\overset{\varphi}{\rightarrow}H^{0}\text{,} \label{EqEmbed2}%
\end{equation}
and
\begin{equation}
G^{0}\overset{\psi}{\rightarrow}\mathbb{R}_{+}\text{.} \label{EqEmbed3}%
\end{equation}
We say that a pair of functions $\varphi,\psi$ as in \emph{(}\ref{EqEmbed2}%
\emph{)} and \emph{(}\ref{EqEmbed3}\emph{)} is \emph{compatible} if the
operator
\begin{equation}
Tu\text{\emph{:}}=u\circ\varphi\label{EqEmbed4}%
\end{equation}
is a well-defined \emph{isometry} mapping $\mathcal{H}_{E}\left(  H\right)  $
\emph{into} $\mathcal{H}_{E}\left(  G\right)  $\emph{;} and if
\begin{equation}
\left(  T\Delta_{H}u\right)  \left(  x\right)  =\psi\left(  x\right)  \left(
\Delta_{G}Tu\right)  \left(  x\right)  \text{.} \label{EqEmbed5}%
\end{equation}

\end{definition}

\begin{definition}
\label{DefEmbed2}\emph{(}with examples!\emph{)} Let $\left(  H,c_{H}\right)  $
be the graph with vertex set $H^{0}$\emph{:}$=\left\{  0\right\}
\cup\mathbb{Z}$, and nearest neighbor edges\emph{;} we set
\begin{equation}
c_{H}\left(  n\right)  \text{\emph{:}}=c_{H}\left(  n-1,n\right)
=2^{n}\text{, for all }n\in\mathbb{Z}_{+}\text{.} \label{EqEmbed6}%
\end{equation}

Let $\left(  G,c_{G}\right)  $ be the graph with vertex set $G^{0}=$ the
dyadic tree \emph{(}see Figure 2\emph{)} extending to $+\infty$ on the right.
\[%
{\parbox[b]{3.7252in}{\begin{center}
\includegraphics[
natheight=4.137800in,
natwidth=3.839800in,
height=4.0116in,
width=3.7252in
]%
{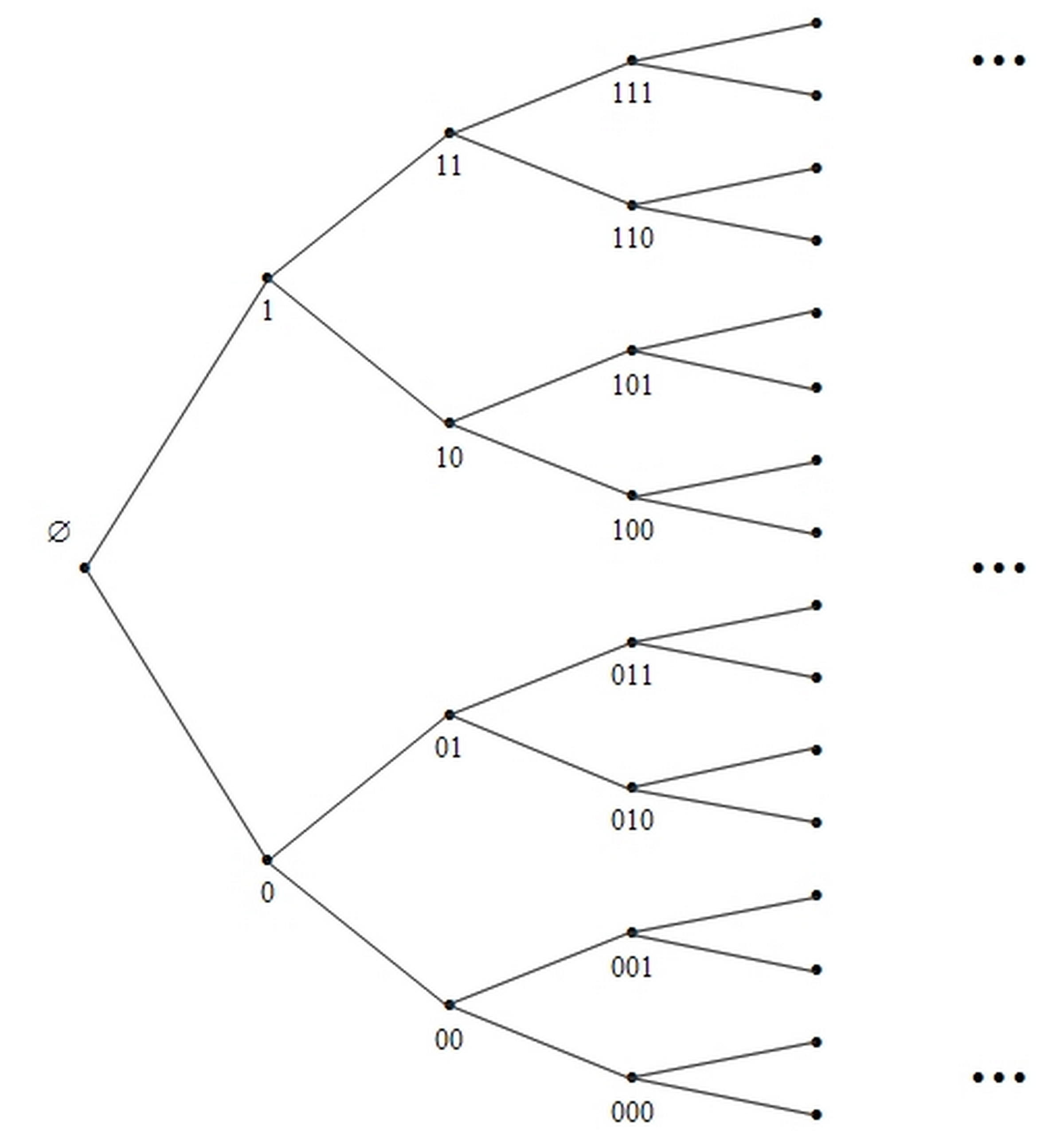}%
\\
\emph{Figure 2. The dyadic tree.}%
\end{center}}}%
\]
Points in $G^{0}$ are finite words in the finite alphabet \emph{(}of
bits\emph{)} $\left\{  0,1\right\}  $. We include the empty word, denoted
$\varnothing$, on the extreme left in Figure 2.

For the edges $G^{1}$, we take the nearest neighbors in $G^{0}$. Hence if
$x\in G^{0}\diagdown\left\{  \varnothing\right\}  $, then $x$ has three
nearest neighbors as follows\emph{:} If $x=\left(  x_{1},x_{2},\ldots
,x_{n}\right)  $, $x_{i}\in\left\{  0,1\right\}  $, then
\begin{equation}
\operatorname*{Nbh}\nolimits_{G}\left(  x\right)  \text{\emph{:}}=\left\{
x^{T},\left(  x0\right)  ,\left(  x1\right)  \right\}  \label{EqEmbed7}%
\end{equation}
where $x^{T}=\left(  x_{1},x_{2}\cdots x_{n-1}\right)  $, $\left(  x0\right)
=\left(  x_{1}\cdots x_{n}0\right)  $, and $\left(  x1\right)  =\left(
x_{1}\cdots x_{n}1\right)  $\emph{;} see also Figure 3.
\[%
{\parbox[b]{1.6837in}{\begin{center}
\includegraphics[
natheight=1.987500in,
natwidth=3.839800in,
height=0.8767in,
width=1.6837in
]%
{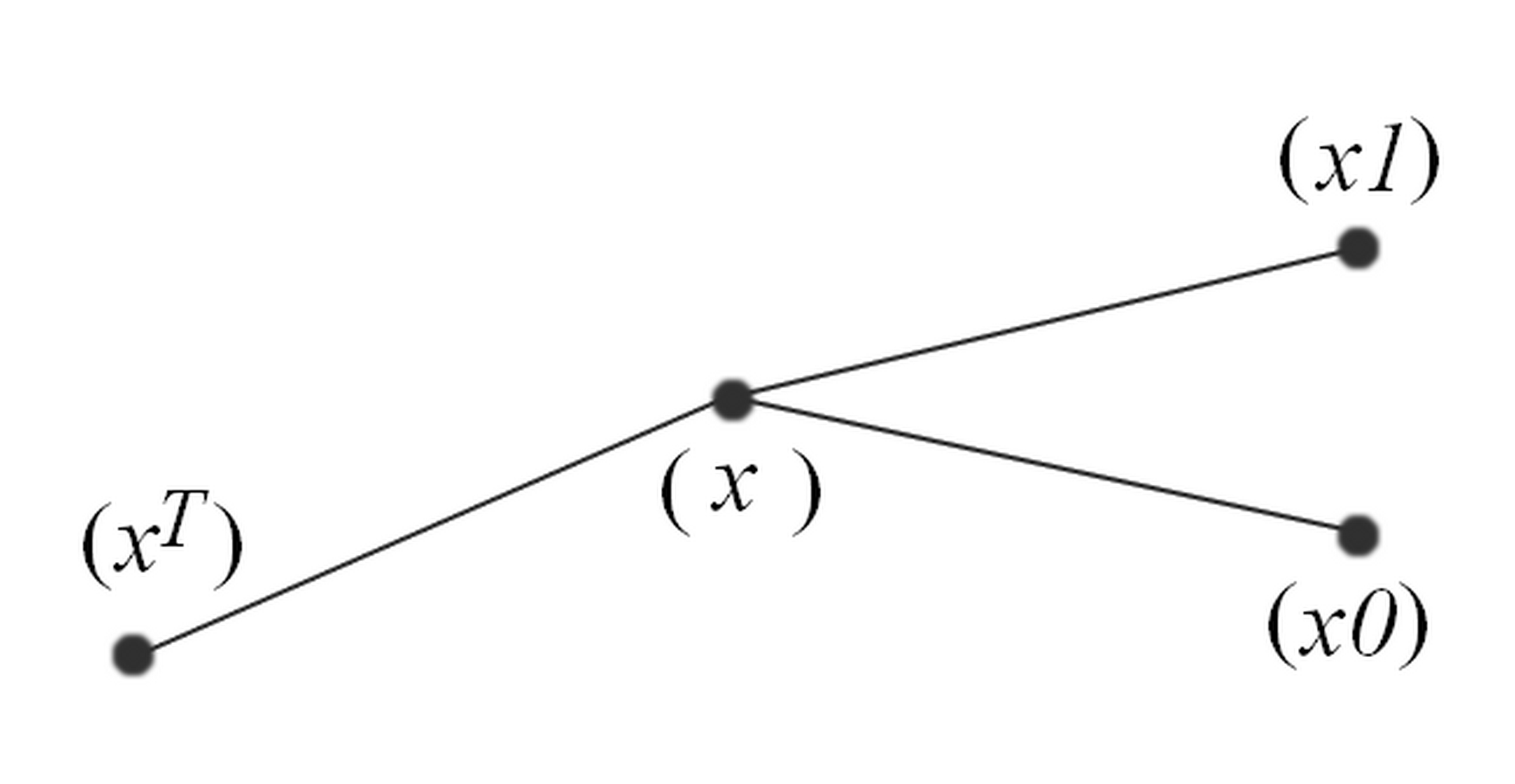}%
\\
\emph{Figure 3. Three nearest Neighbors.}%
\end{center}}}%
\]

\end{definition}

The remaining case is
\begin{equation}
\operatorname*{Nbh}\nolimits_{G}\left(  \varnothing\right)  =\left\{
0,1\right\}  \text{.} \label{EqEmbed8}%
\end{equation}

The reader may check by computation that the following two functions
$\varphi:G^{0}\rightarrow H^{0}$, and $\psi:G^{0}\rightarrow\mathbb{R}_{+}$
constitute a compatible pair; i.e., that if $T$ is defined as in
(\ref{EqEmbed4}) then it is \textit{isometric} and $\psi$-intertwining, see
(\ref{EqEmbed5}). We set
\begin{align}
\varphi\left(  x_{1}x_{2}\cdots x_{n}\right)   &  =n\text{ if }x\not =%
\varnothing\text{,}\nonumber\\
\varphi\left(  \varnothing\right)   &  =0,\text{and }\label{EqEmbed9}\\
\psi\left(  x\right)   &  :=2^{\varphi\left(  x\right)  },~x\in G^{0}%
\text{.}\nonumber
\end{align}

We now turn to some applications dealing with harmonic functions of finite
energy, and monopoles.

\begin{definition}
\label{DefEmbed3}Let $\left(  G,c\right)  $ be a weighted graph with
conductance function $c$\emph{;} and let $\Delta$ and $\mathcal{H}_{E}$ be the
corresponding Laplace operator, and energy Hilbert space, respectively. Pick
$o\in G^{0}$\emph{;} and set
\begin{equation}
\operatorname*{Mono}\left(  G\right)  \text{\emph{:}}=\left\{  w\in
\mathcal{H}_{E}|\Delta^{\ast}w=-\delta_{0}\right\}  \label{EqEmbed10}%
\end{equation}
and
\begin{equation}
\operatorname*{Harm}\left(  G\right)  \text{\emph{:}}=\left\{  u\in
\mathcal{H}_{E}|\Delta^{\ast}u=0\right\}  \text{.} \label{EqEmbed11}%
\end{equation}
Here \textquotedblleft Mono\textquotedblright\ is short for \textquotedblleft
monopole\textquotedblright\emph{;} and \textquotedblleft
Harm\textquotedblright\ is short for \textquotedblleft
harmonic.\textquotedblright
\end{definition}

\begin{proposition}
\label{PropEmbed1}Let $\left(  G,c_{G}\right)  $ and $\left(  H,c_{H}\right)
$ be weighted graphs, and let $\varphi,\psi$ be a pair of functions satisfying
the conditions in \emph{(}\ref{EqEmbed2}\emph{)} and \emph{(}\ref{EqEmbed3}%
\emph{)}. Assume the two functions are compatible such that $T$ in
\emph{(}\ref{EqEmbed4}\emph{)} is isometric $\mathcal{H}_{E}\left(  H\right)
\hookrightarrow\mathcal{H}_{E}\left(  G\right)  $, and satisfies the
intertwining property \emph{(}\ref{EqEmbed5}\emph{)}.

Then $T$ maps $\operatorname*{Harm}\left(  H\right)  $ into
$\operatorname*{Harm}\left(  G\right)  $, and $\operatorname*{Mono}\left(
H\right)  $ into $\operatorname*{Mono}\left(  G\right)  $.
\end{proposition}

\begin{proof}
The two assertions in the proposition are the following inclusions:
\begin{equation}
T\left(  \operatorname*{Harm}\left(  H\right)  \right)  \subseteq
\operatorname*{Harm}\left(  G\right)  \text{,} \label{EqEmbed12}%
\end{equation}
and
\begin{equation}
T\left(  \operatorname*{Mono}\left(  H\right)  \right)  \subseteq
\operatorname*{Mono}\left(  G\right)  \text{.} \label{EqEmbed13}%
\end{equation}

Now take $u\in\operatorname*{Harm}\left(  H\right)  $; i.e., $\Delta_{H}u=0$.
An application of (\ref{EqEmbed5}) yields:
\[
\psi\left(  x\right)  \left(  \Delta_{G}Tu\right)  \left(  x\right)  =\left(
T\Delta_{H}u\right)  \left(  x\right)  =0\text{.}%
\]
Since $\Delta_{H}u\equiv0$ and $\psi\left(  x\right)  \not =0$, we conclude
that $\Delta_{G}Tu=0$, i.e., that
\[
Tu\in\operatorname*{Harm}\left(  G\right)  \text{.}%
\]

For the second inclusion, we pick points $o_{H}$ and $o_{G}$ in the respective
vertex sets such that $\varphi\left(  o_{G}\right)  =o_{H}$.

If $w\in\operatorname*{Mono}\left(  H\right)  $, then a second application of
(\ref{EqEmbed5}) shows that
\[
\Delta_{G}Tw=-\psi\left(  o_{G}\right)  \delta_{o_{G}}\text{;}%
\]
and so $Tw\in\operatorname*{Mono}\left(  G\right)  $.
\end{proof}

\begin{corollary}
\label{CorEmbed1}For the dyadic tree $G$ in Figures 2 and 3 with constant
conductance, we have
\[
\operatorname*{Harm}\left(  G\right)  \not =0\text{, and }\operatorname*{Mono}%
\left(  G\right)  \not =0\text{.}%
\]

\end{corollary}

\begin{proof}
A direct application of the proposition.
\end{proof}

\section{Two Hilbert Spaces\label{Hilbert}}

Since the summations from (\ref{EqSpace7}) in Proposition \ref{PropSpace1},
and on the right-hand side in Theorem \ref{TheoSpace1}, involve $\ell^{2}%
$-considerations, we now turn to a comparison of the two Hilbert spaces
$\ell^{2}$ and $\mathcal{H}_{E}$.

Recall the data $\left(  G,c\right)  $, $\Delta$ and $\mathcal{H}_{E}$ are as
described above; in particular
\begin{equation}
c:G^{1}\rightarrow\mathbb{R}_{+} \label{EqHilbert1}%
\end{equation}
is a fixed function, and the conditions listed in Definition \ref{DefFrame1}
are assumed to hold. Note further that by Definition \ref{DefFrame1}(c), the
norm, and the inner product, $\left\Vert \cdot\right\Vert _{\mathcal{H}_{E}}$
and $\left\langle \cdot,\cdot\right\rangle _{E}$ (\textquotedblleft
E\textquotedblright\ for energy) are weighted with the fixed function $c$ in
(\ref{EqHilbert1}); and the definitions in the energy Hilbert space
$\mathcal{H}_{E}$ involve differences $u\left(  x\right)  -u\left(  y\right)
$ whenever $\left(  x,y\right)  \in G^{1}$.

In contrast the norm $\left\Vert \cdot\right\Vert _{2}$ and inner product
$\left\langle \cdot,\cdot\right\rangle _{\ell^{2}}$ are unweighted; in fact we
have:
\begin{equation}
\left\Vert u\right\Vert _{2}^{2}=\sum_{x\in G^{0}}\left\vert u\left(
x\right)  \right\vert ^{2}\text{;} \label{EqHilbert2}%
\end{equation}
and
\begin{equation}
\left\langle u,v\right\rangle _{\ell^{2}}=\sum_{x\in G^{0}}\overline{u\left(
x\right)  \,}v\left(  x\right)  \text{, for all }u,v\in\ell^{2}=\ell
^{2}\left(  G^{0}\right)  \text{.} \label{EqHilbert3}%
\end{equation}

It is immediate that
\begin{equation}
\mathcal{D}_{2}\text{:}=\operatorname*{span}\left\{  \delta_{x}\,|\,x\in
G^{0}\right\}  \label{EqHilbert4}%
\end{equation}
is a dense linear subspace in $\ell^{2}$, and further that the Laplace
operator $\Delta$ from (\ref{EqFrame6}) satisfies
\begin{equation}
\left\langle u,\Delta u\right\rangle _{\ell^{2}}\geq0\text{ for all }%
u\in\mathcal{D}_{2}\text{.} \label{EqHilbert5}%
\end{equation}
In other words, it defines a semibounded Hermitian operator whith dense
domains in $\ell^{2}$. Moverover, by \cite{JoPe08}, this operator $\Delta$
(with domain $\mathcal{D}_{2}$) is essentially selfadjoint in $\ell^{2}$. This
means that its graph closure in $\ell^{2}\times\ell^{2}$ is a selfadjoint
operator in $\ell^{2}$.

However, as we show in sections \ref{Compute}--\ref{Poly} above, $\Delta$ is
not essentially selfadjoint when viewed as an unbounded operator with dense
domain in $\mathcal{H}_{E}$; see especially Theorem \ref{TheoSpace1}.

Our terminology for the operator $\Delta$ in equation (\ref{EqFrame6}) will
involve some ambiguity. In fact, $\Delta$ will simultaneously be viewed as a
densely defined operator in $\ell^{2}$, and in $\mathcal{H}_{E}$; the Hilbert
spaces are different, and the respective dense domains are different as well.
Each of the two operators will be closed; the first in the graph norm in
$\ell^{2}\times\ell^{2}$, and the second in $\mathcal{H}_{E}\times
\mathcal{H}_{E}$. The first of the two closed operators is selfadjoint, while
the second generally is not. What is worse, the two Hilbert spaces $\ell^{2}$
and $\mathcal{H}_{E}$ do not lend themselves to a direct comparison.

We shall need the following:

\begin{lemma}
\label{LemHilbert1} ~

\begin{enumerate}
\item[(a)] Let $\Delta_{2}$ denote the selfadjoint operator in $\ell^{2}$, and
let $\operatorname*{dom}\left(  \Delta_{2}\right)  $ be its dense domain. Then
the sum-operator $I+\Delta_{2}$ is invertible, and the following hold\emph{:}
\begin{equation}
\left(  I+\Delta_{2}\right)  \left(  \operatorname*{dom}\left(  \Delta
_{2}\right)  \right)  =\ell^{2}\text{,} \label{EqHilbert6}%
\end{equation}
and
\begin{equation}
\left(  I+\Delta_{2}\right)  ^{-1}\ell^{2}=\operatorname*{dom}\left(
\Delta_{2}\right)  \text{.} \label{EqHilbert7}%
\end{equation}

\item[(b)] Let $\Delta_{E}$ denote the closed operator version of $\Delta$
with dense domain $\operatorname*{dom}\left(  \Delta_{E}\right)  $ in
$\mathcal{H}_{E}$. Then the following hold\emph{:}
\begin{equation}
\left(  I+\Delta_{E}\right)  \left(  \operatorname*{dom}\left(  \Delta
_{E}\right)  \right)  =\mathcal{H}_{E}\ominus\operatorname*{Def}\text{,}
\label{EqHilbert8}%
\end{equation}
and $\left(  I+\Delta_{E}\right)  ^{-1}$ is well-defined on this closed subspace.

\item[(c)] The operator $\left(  I+\Delta_{2}\right)  ^{-1}$ is contractive
from $\ell^{2}$ to $\ell^{2}$\emph{;} i.e.,
\begin{equation}
\left\Vert \left(  I+\Delta_{2}\right)  ^{-1}u\right\Vert _{2}\leq\left\Vert
u\right\Vert _{2}\text{ for all }u\in\ell^{2}\text{.} \label{EqHilbert9}%
\end{equation}

\item[(d)] The operator $\left(  I+\Delta_{E}\right)  ^{-1}$ is contractive
from its domain \emph{(}\ref{EqHilbert8}\emph{)} into $\mathcal{H}_{E}$,
i.e.,
\begin{equation}
\left\Vert \left(  I+\Delta_{E}\right)  ^{-1}u\right\Vert _{E}\leq\left\Vert
u\right\Vert _{E} \label{EqHilbert10}%
\end{equation}
holds for all $u$ of the form $u=v+\Delta_{E}v$.
\end{enumerate}
\end{lemma}

\begin{proof}
The verification of the conclusions listed in (a)--(d) follow by combining the
properties of $\Delta$ which we derived in sections \ref{Compute} and
\ref{Random}--\ref{Space} above.
\end{proof}

\begin{theorem}
\label{TheoHilbert1}Let $\ell^{2},\mathcal{H}_{E},$ and $\Delta$ be as
specified in the lemma where $\Delta$ will have the meaning which is dictated
by the contexts of the respective Hilbert spaces.

\begin{enumerate}
\item[(a)] Then we have the following containment
\begin{equation}
\left(  I+\Delta_{2}\right)  ^{-1}\ell^{2}=\left\{  u\in\ell^{2}\,|\,\Delta
u\in\ell^{2}\right\}  \subset\mathcal{H}_{E}\text{.} \label{EqHilbert11}%
\end{equation}

\item[(b)] If $u\in\ell^{2}$ and $\Delta u\in\ell^{2}$, then
\begin{equation}
\mathcal{E}\left(  u\right)  =\left\Vert u\right\Vert _{\mathcal{H}_{E}}%
^{2}=\sum_{x\in G^{0}}\overline{u\left(  x\right)  }\left(  \Delta v\right)
\left(  x\right)  \text{.} \label{EqHilbert12}%
\end{equation}

\end{enumerate}
\end{theorem}

\begin{proof}
Let $u\in\left(  I+\Delta_{2}\right)  ^{-1}\ell^{2}$ and pick $v\in\ell^{2}$
such that $u+\Delta u=v$. When the inner product of $\ell^{2}$ is considered,
this means that $u$ is in the domain of $\Delta_{2}^{\ast}$, and $\Delta
_{2}^{\ast}u=v-u\in\ell^{2}$. But $\Delta_{2}$ is essentially sefadjoint by
the lemma, so $\Delta_{2}^{\ast}$ is the closure of $\Delta_{2}$. Moreover
$\Delta^{\ast}u$ is given by the formula (\ref{EqFrame6}) in section
\ref{Frame}.

Conversely, suppose $u$ and $\Delta u$ are in $\ell^{2}$; then set
$v$:$=u+\Delta u$, and note that $\left(  I+\Delta_{2}\right)  ^{-1}v=u$.

Our aim is now to show that the double-summation (\ref{EqFrame4}) in
Definition \ref{DefFrame1} is finite; and so $u\in\mathcal{H}_{E}$. The
conditions on the function $c$ are as specified in Definition \ref{DefFrame1}(a).

When the double-summation is carried out, we note that Fubini applies; we get
\begin{align*}
&  \frac{1}{2}\underset{%
\genfrac{}{}{0pt}{}{\text{all }x,y}{\text{s.t. }x\sim y}%
}{\sum\sum}c\left(  x,y\right)  \left\vert u\left(  x\right)  -u\left(
y\right)  \right\vert ^{2}\\
&  =\sum_{x}\left(  c\left(  x\right)  \left\vert u\left(  x\right)
\right\vert ^{2}-\sum_{y\sim x}\left(  c\left(  x,y\right)  \overline{u\left(
x\right)  }\,u\left(  y\right)  \right)  \right) \\
&  =\sum_{x}\overline{u\left(  x\right)  }\left(  c\left(  x\right)  u\left(
x\right)  -\sum_{y\sim x}c\left(  x,y\right)  u\left(  y\right)  \right) \\
&  =\sum_{x}\overline{u\left(  x\right)  }\left(  \Delta u\right)  \left(
x\right)  \text{.}%
\end{align*}

\end{proof}

\begin{corollary}
\label{CorHilbert1}Let $\ell^{2},\mathcal{H}_{E}$, and $\Delta$ be as
specified in the theorem, and let $x\in G^{0}$ be given.

\begin{enumerate}
\item[(a)] Then the space of solutions $u\in\mathcal{H}_{E}$ to the equation
\begin{equation}
\Delta u=-u\text{ in }G^{0}\diagdown\left\{  x\right\}  \text{,}
\label{EqHilbert13}%
\end{equation}
is one-dimensional. \emph{(}Note that \emph{(}\ref{EqHilbert13}\emph{)} is the
deficiency equation in the \emph{punctured} vertex set $G^{0}\diagdown\left\{
x\right\}  $.\emph{)}

\item[(b)] The solution space in $\mathcal{H}_{E}$ from \emph{(}%
\ref{EqHilbert13}\emph{)} is spanned by $\left(  I+\Delta\right)  ^{-1}%
\delta_{x}$.
\end{enumerate}
\end{corollary}

\begin{proof}
The essential point is that if
\begin{equation}
u_{x}=\left(  I+\Delta_{2}\right)  ^{-1}\left(  \delta_{x}\right)
\label{EqHilbert14}%
\end{equation}
is computed first in $\ell^{2}$, then the two vectors $u_{x}$ and $\Delta
u_{x}=\delta_{x}-u_{x}$ are both in $\ell^{2}$. As a result (see Theorem
\ref{TheoHilbert1}(a)) we conclude that $u_{x}$ is in $\mathcal{H}_{E}$. It
follows that
\begin{equation}
u_{x}+\Delta u_{x}=\delta_{x} \label{EqHilbert15}%
\end{equation}
holds pointwise on $G^{0}$, and so
\[
u_{x}+\Delta u_{x}=0\text{ on }G^{0}\diagdown\left\{  x\right\}  \text{.}%
\]

Conversely, if $u\in\mathcal{H}_{E}$ satisfies (\ref{EqHilbert13}) then it
follows that for some constant $C$ we have $u+\Delta u=C\delta_{x}$; and
therefore
\begin{equation}
u=C\left(  I+\Delta\right)  ^{-1}\delta_{x}=Cu_{x}\text{.} \label{EqHilbert16}%
\end{equation}

This proves that the solution space in (\ref{EqHilbert13}) is one-dimensional
and spanned by the single function $u_{x}\left(  \in\mathcal{H}_{E}\right)  $
in (\ref{EqHilbert14}).\bigskip
\end{proof}

\begin{center}
\textsc{Acknowledgment}
\end{center}

The author is happy to thank his colleagues (for suggestions) in the math
physics and operator theory seminars at the University of Iowa, where he
benefitted from numerous helpful discussions of various aspects of the results
in the paper, and their applications. And he thanks Doug Slauson for
typesetting the paper.

This work supported in part by the US National Science Foundation.

\newif\ifabfull\abfulltrue
\providecommand{\bysame}{\leavevmode\hbox to3em{\hrulefill}\thinspace}
\providecommand{\MR}{\relax\ifhmode\unskip\space\fi MR }
\providecommand{\MRhref}[2]{%
  \href{http://www.ams.org/mathscinet-getitem?mr=#1}{#2}
}
\providecommand{\href}[2]{#2}

\end{document}